\documentclass[preprintnumbers,article,amsmath,amssymb,floatfix,10pt,prd,twocolumn,
superscriptaddress,nofootinbib]{revtex4-2}
\usepackage{bm}
\usepackage{amsfonts}
\usepackage{latexsym}
\usepackage[latin1]{inputenc}
\usepackage{graphicx}
\usepackage{amsmath}
\usepackage{bigints} 
\usepackage{palatino}
\usepackage{mathrsfs} 
\usepackage{mathpazo}
\usepackage{textcomp}
\usepackage{float}
\usepackage{booktabs}
\usepackage{dcolumn}
\usepackage{ragged2e}
\usepackage{hyperref}
\usepackage{relsize} 
\hypersetup{colorlinks,citecolor=blue}
\hypersetup{colorlinks=true,linkcolor=red,filecolor=magenta,    urlcolor=blue}
\usepackage{amsmath}
\usepackage{xcolor}
\usepackage{orcidlink}
\usepackage{epsfig}
\usepackage{subfigure}
\usepackage{commath}

\usepackage{titlesec}
\titleclass{\subsubsubsection}{straight}[\subsection]
\newcounter{subsubsubsection}[subsubsection]
\renewcommand\thesubsubsubsection{\thesubsubsection.\arabic{subsubsubsection}}
\titleformat{\subsubsubsection}
  {\normalfont\normalsize\bfseries}{\thesubsubsubsection}{1em}{}
\titlespacing*{\subsubsubsection}
{0pt}{3.25ex plus 1ex minus .2ex}{1.5ex plus .2ex}
\makeatletter
\renewcommand\paragraph{\@startsection{paragraph}{5}{\z@}%
  {3.25ex \@plus1ex \@minus.2ex}%
  {-1em}%
  {\normalfont\normalsize\bfseries}}
\renewcommand\subparagraph{\@startsection{subparagraph}{6}{\parindent}%
  {3.25ex \@plus1ex \@minus .2ex}%
  {-1em}%
  {\normalfont\normalsize\bfseries}}
\def\toclevel@subsubsubsection{4}
\def\toclevel@paragraph{5}
\def\toclevel@paragraph{6}
\def\l@subsubsubsection{\@dottedtocline{4}{7em}{4em}}
\def\l@paragraph{\@dottedtocline{5}{10em}{5em}}
\def\l@subparagraph{\@dottedtocline{6}{14em}{6em}}
\makeatother
\DeclareFontFamily{U}{BOONDOX-calo}{\skewchar\font=45 }
\DeclareFontShape{U}{BOONDOX-calo}{m}{n}{
  <-> s*[1.05] BOONDOX-r-calo}{}
\DeclareFontShape{U}{BOONDOX-calo}{b}{n}{
  <-> s*[1.05] BOONDOX-b-calo}{}
\DeclareMathAlphabet{\mathcalboondox}{U}{BOONDOX-calo}{m}{n}
\SetMathAlphabet{\mathcalboondox}{bold}{U}{BOONDOX-calo}{b}{n}
\DeclareMathAlphabet{\mathbcalboondox}{U}{BOONDOX-calo}{b}{n}

\def\jnl@style{\it}
\def\aaref@jnl#1{{\jnl@style#1}}

\def\aaref@jnl#1{{\jnl@style#1}}

\def\aj{\aaref@jnl{AJ}}                   
\def\apj{\aaref@jnl{ApJ}}                 
\def\apjl{\aaref@jnl{ApJ}}                
\def\apjs{\aaref@jnl{ApJS}}               
\def\apss{\aaref@jnl{Ap\&SS}}             
\def\aap{\aaref@jnl{A\&A}}                
\def\aapr{\aaref@jnl{A\&A~Rev.}}          
\def\aaps{\aaref@jnl{A\&AS}}              
\def\mnras{\aaref@jnl{Mon.~Not.~Roy.~Astron.~Soc.}}             
\def\prd{\aaref@jnl{Phys.~Rev.~D}}        
\def\prc{\aaref@jnl{Phys.~Rev.~C}}  
\def\prl{\aaref@jnl{Phys.~Rev.~Lett.}}    
\def\qjras{\aaref@jnl{QJRAS}}             
\def\skytel{\aaref@jnl{S\&T}}             
\def\ssr{\aaref@jnl{Space~Sci.~Rev.}}     
\def\zap{\aaref@jnl{ZAp}}                 
\def\nat{\aaref@jnl{Nature}}              
\def\aplett{\aaref@jnl{Astrophys.~Lett.}} 
\def\apspr{\aaref@jnl{Astrophys.~Space~Phys.~Res.}} 
\def\physrep{\aaref@jnl{Phys.~Rep.}}      
\def\physscr{\aaref@jnl{Phys.~Scr}}       
\def\commat{\aaref@jnl{Comm.~Math.~Phys.}}              
\def\science{\aaref@jnl{Science}}               
\def\cqg{\aaref@jnl{Classical Quant.~Grav.}}            
\def\jpcs{\aaref@jnl{JPCS}}                                     
\def\ijmpd{\aaref@jnl{Int.~J.~Mod.~Phys.~D}}                    
\def\grg{\aaref@jnl{Gen.~Relat.~Gravit.}}               
\def\rpp{\aaref@jnl{Rep.~Prog.~Phys.}}          
\def\npa{\aaref@jnl{Nucl.~Phys.~A}}        
\def\lrr{\aaref@jnl{Living Rev.~Rel.}}                   
\def\jcap{\aaref@jnl{J.~Cosmology Astropart.~Phys.}}    
\def\rmp{\aaref@jnl{Rev.~Mod.~Phys.}}   
\def\epjc{\aaref@jnl{Eur.~Phys.~J.~C}} 
\def\plb{\aaref@jnl{~Phy.~Lett.~B}} 
\def\mpla{\aaref@jnl{Mod.~Phy.~Lett.~A}} 
\def\arxiv{\aaref@jnl{arxiv.org}}

\allowdisplaybreaks[1]

\begin{document}
\color{black} 
\title{Wormhole formations in the galactic halos supported by dark matter models and global monopole charge within $f(Q)$ gravity}

\author{Moreshwar Tayde\orcidlink{0000-0002-3110-3411}}
\email{moreshwartayde@gmail.com}
\affiliation{Department of Mathematics, Birla Institute of Technology and
Science-Pilani,\\ Hyderabad Campus, Hyderabad-500078, India.}

\author{P.K. Sahoo\orcidlink{0000-0003-2130-8832}}
\email{pksahoo@hyderabad.bits-pilani.ac.in}
\affiliation{Department of Mathematics, Birla Institute of Technology and
Science-Pilani,\\ Hyderabad Campus, Hyderabad-500078, India.}

%
\date{\today}

\begin{abstract}
This paper discusses the possibility of traversable wormholes in the galactic region supported by dark matter (DM) models and global monopole charge in the context of $f(Q)$ gravity. To understand the features of the wormholes, we comprehensively studied wormhole solutions with various redshift functions under different $f(Q)$ models. We obtained wormhole shape functions for Pseudo Isothermal (PI) and Navarro-Frenk-White (NFW) DM profiles under linear $f(Q)$ gravity. In contrast, we employed an embedding class I approach for the non-linear $f(Q)$ models to investigate wormholes. We noticed that our obtained shape functions satisfy the flare-out conditions under an asymptotic background for each DM profile. Moreover, we checked the energy conditions at the wormhole throat with a radius $r_0$ and noticed the influences of the global monopole's parameter $\eta$ in the violation of energy conditions, especially null energy conditions. Further, for the non-linear case, we observed that wormhole solutions could not exist for $f(Q)=Q+mQ^n$, $f(Q)=Q+\frac{\beta}{Q}$, and $f(Q)=\alpha_1+\beta_1 \log(Q)$ under embedding class I approach. Finally, we study the amount of exotic matter via the volume integral quantifier technique for the linear $f(Q)$ model, and we confirm that a small amount of exotic matter is required to sustain the traversable wormholes. 
\end{abstract}

\maketitle



\section{Introduction}\label{sec1}
Wormholes represent solutions of the Einstein field equations within the framework of the general theory of relativity, manifesting as tunnel-like structures capable of linking disparate regions of space-time within our Universe or even different universes \cite{L. Flamm, A. Einstein}. Notably, Einstein and Rosen \cite{A. Einstein} postulated a geometric interpretation for elementary particles through the Einstein-Rosen bridge (ERB), though this model was ultimately unsuccessful \cite{J. Wheeler, H. G. Ellis, R. W. Fuller}. Traversable wormholes have since sparked widespread interest in the field of science. The exploration of traversable wormholes gained momentum with seminal contributions such as Ellis' and Bronnikov's investigations into traversable wormholes featuring a phantom scalar field \cite{H. G. Ellis, K. A. Bronnikov}. Subsequent studies, including those by Clement and the groundbreaking work of Morris and Thorne, have delved into various wormhole models \cite{G. Clement,  M. S. Morris}.\\
\indent A defining characteristic of traversable wormhole geometry is the necessity for exotic matter concentrated at the wormhole throat to maintain its spatial openness, thereby violating energy conditions such as the null energy condition (NEC) \cite{M. Visser 1}. A notable development in the study of traversable wormholes is the work by Kanti et al., who devised wormhole constructions within the framework of quadratic gravitational theories \cite{P. Kanti 1, P. Kanti 2}. This innovative approach reveals a fascinating concept wherein gravity plays a pivotal role in maintaining the openness of the wormhole throat, obviating the necessity for exotic matter. Further, Kuhfittig \cite{P. K. H. Kuhfittig} investigated rotating axially symmetric wormholes by incorporating time-dependent angular velocity, aiming to generalize static and spherically symmetric traversable wormholes. Barros and Lobo \cite{B. J. Barros} derived wormhole solutions utilizing three form fields and scrutinized the compliance with weak and null energy conditions. Also, A. \"Ovg\"un \cite{B9} constructed a rotating thin-shell wormhole using a Myers-Perry black hole in five dimensions, using the Darmois-Israel junction conditions and found that the exotic matter is required at the throat of the wormhole to keep it stable.\\
\indent Wormhole geometries have received significant attention not just in modified gravity theories but also in higher-dimensional gravitational theories \cite{Mak, Zangeneh, Kar2, Ziaie} and Kaluza-Klein gravity \cite{Singleton, Leon}. These theoretical frameworks have different benefits, such as eliminating the necessity for nonstandard fluids, a major motivator for extensive study in modified gravity theories. Furthermore, changes to Einstein's gravity introduced additional degrees of freedom inside the gravitational sector, opening up new options for addressing difficulties such as dark energy and dark matter. The evolution of wormhole geometries has been studied in modified gravity theories, including $f(R)$ \cite{Pavlovic, Idris}. Mazharimousavi and Halilsoy \cite{Halilsoy} investigated how wormholes manifest in vacuum and non-vacuum environments. Their work has resulted in stable wormhole geometries through the $f(R)$ model, focusing on polynomial evolution. Also, authors \cite{V. De Falco} focused on $f(R)$ metric, $f(T)$ teleparallel, and $f(Q)$ symmetric teleparallel models to explore static and spherically symmetric wormhole solutions. Furthermore, recent literature presents intriguing examinations of wormhole geometry within various modified gravity theories \cite{Golchin, Ahmad, Bhatti, Chanda, Rosa, Tayde 1, Tayde 2, C. G. Bohmer 2, Rani, Tayde 3, Sakalh, A6}.\\
\indent Wormholes serve as valuable testbeds for various theories of gravity \cite{S. Capozziello 1, S. Capozziello 2}. Among these, a noteworthy contender is $f(Q)$ gravity, introduced by Jimenez et al. \cite{J. B. Jimenez}, wherein the gravitational interaction is governed by the nonmetricity $Q$. Over the past years, $f(Q)$ gravity has undergone extensive observational scrutiny, with Lazkoz et al. \cite{R. Lazkoz} proposing intriguing constraints on this theory. Furthermore, Mandal et al. \cite{S. Mandal} successfully demonstrated the viability of $f(Q)$ gravity models concerning energy conditions. Moreover, Hassan et al. \cite{Zinnat 2} have systematically analyzed the impact of the Generalized Uncertainty Principle (GUP) on the Casimir wormhole space-time within the framework of $f(Q)$ gravity. Significantly, numerous captivating astrophysical investigations have been undertaken within the $f(Q)$ gravity and its extended gravity theory. Interested readers can delve deeper into this rich literature, with references \cite{L. Heisenberg, F. Parsaei, O. Sokoliuk 1, O. Sokoliuk 3, S. Pradhan, Debasmita} providing valuable sources for additional exploration.\\
\indent Recent findings concerning the Universe have unveiled that visible matter, including planets, stars, and other observable entities, constitutes approximately 5\% of the total matter and energy content. The remaining portion is primarily governed by dark matter (DM) and dark energy, with DM accounting for about 27\% of the total composition. DM is recognized as a pivotal element in the formation and progression of galaxies on a galactic scale \cite{S. Trujillo-Gomez}. Given the widespread distribution of DM halos and galaxies, it becomes imperative to investigate the potential formation of traversable wormholes within these cosmic structures \cite{F. Rahaman 1, F. Rahaman 2}. Previous research, such as that presented in \cite{F. Rahaman 1}, has explored the feasibility of traversable wormhole formation in the outer regions of galactic halos using the Navarro-Frenk-White (NFW) profile. Moreover, Xu et al. \cite{Z. Xu} delve into the feasibility of traversable wormhole formation within the enigmatic realms of DM halos with the NFW, Thomas-Fermi (TF), and Pseudo Isothermal (PI) matter density profiles. Furthermore, \cite{Rahaman11} introduced a novel interpretation of traversable wormholes grounded in Einstein's field equations. This model was bolstered by the incorporation of Einasto DM density profiles and global monopole charges, alongside accounting for semiclassical effects within the local Universe, particularly within galactic halos.\\
\indent It is well known that, unlike electric charge, magnetic monopole can not exist for Maxwell's equations in $\mathbb{R}^4$ (for simply connected manifolds). However, Dirac \cite{P. A. M. Dirac} first pointed out that if one introduces a quantization condition with a magnetic monopole (changing the topology of the manifold by taking a ``Dirac string" out), then one can show that the electric charges are quantized. Even though it was a great and natural way to explain the discreteness of the electric charge, it required non-singular transformation (like taking the ``Dirac string" out), it was later found out by  G't Hooft \cite{G.'t Hooft} and Polyakov \cite{A. M. Polyakov} that for non-abelian gauge theories, there can be monopole. Arising from spontaneous symmetry breaking and ground state having non-trivial topology. Also, unlike the Dirac monopole, there is a singular transformation required to get these monopoles, so it was conjectured in their paper that such a monopole could naturally arise during the early Universe when the GUT phase transitions were taking place. It was later shown by Barriola and Valenkin \cite{M. Barriola} that scalar field having $SO(3)$ symmetry near Schwarchuild metric could go through a spontaneous symmetry-breaking mechanism and give monopole. In this paper, we study wormhole geometry under the influence of monopole and galactic halo in $f(Q)$ gravity. Even though the study of wormholes in the context of monopole \cite{Tayde 6, Rahaman11, S. Sarkar, P. Das} and galactic halo \cite{Z. Xu, Rahaman11, G. Mustafa 2, Tayde 5, B7} have been studied before. Here, we give a more general picture with $f(Q)$ gravity.\\
\indent The structure of this paper unfolds as follows: Section \ref{sec2} delineates the criteria defining a traversable wormhole imbued with a Global Monopole alongside the formalism elucidating $f(Q)$ gravity. Section \ref{sec3} is dedicated to deriving the field equations within a linear model framework. Following this, in Section \ref{sec4}, we introduce two distinctive DM profiles and meticulously examine the requisite conditions for the existence of a traversable wormhole while also scrutinizing the energy conditions under different redshift functions. Section \ref{sec5} expands our analysis to incorporate the non-linear formulation of $f(Q)$ concerning DM halo profiles. Section \ref{sec6} is dedicated to utilizing a volume integral quantifier parameter to evaluate the quantity of exotic matter necessary. Finally, in the concluding Section \ref{sec7}, we consolidate our findings and engage in a comprehensive discussion concerning the implications of the results derived from this study. 

\section{Basic Criteria of a traversable wormhole with Global Monopole and formalism of $f(Q)$ gravity}
\label{sec2}
The wormhole metric outlined by Morris \cite{M. S. Morris} and Visser \cite{M. Visser 1} in the Schwarzschild coordinates $(t,r,\theta,\Phi)$, is expressed as follows:
\begin{equation} \label{11}
ds^{2}=e^{\nu(r)}dt^{2}-e^{\lambda(r)} dr^{2}-r^{2} d\theta^{2}-r^{2}\sin^{2}\theta d\Phi^{2}.
\end{equation}
Here, the metric components, $\nu(r) = 2\phi(r)$ and $e^{\lambda(r)} = \left(\frac{r - b(r)}{r}\right)^{-1}$, are functions solely dependent on the radial coordinate. In this context, $b(r)$ serves as the shape function that characterizes the geometry of the wormhole, while $\phi(r)$ is the redshift function linked to the gravitational redshift effect. For a wormhole to be traversable, the shape function $b(r)$ must meet the flaring-out condition, expressed by $(b - b'r)/b^2 > 0$ \cite{M. S. Morris}. At the throat of the wormhole, where $b(r_0) = r_0$, the requirement $b^{,\prime}(r_0) < 1$ must hold, with $r_0$ representing the radius of the throat. Moreover, the condition of asymptotic flatness dictates that as $r \rightarrow \infty$, the ratio $\frac{b(r)}{r}$ approaches zero. To prevent the formation of an event horizon, $\phi(r)$ must remain finite throughout. These criteria ensure the potential presence of exotic matter at the wormhole throat within the framework of Einstein's General Relativity.

The action for symmetric teleparallel gravity incorporating a global monopole charge, derived from a four-dimensional framework without including the cosmological constant and with a minimally coupled triplet scalar field (with $\hbox {c} = \hbox {G} = 1$), can be formulated as follows:
\begin{equation}\label{1}
\mathcal{S}=\int\frac{1}{16\pi}\,f(Q)\sqrt{-g}\,d^4x+\int (\mathcal{L}_m+\mathcal{L})\,\sqrt{-g}\,d^4x\,.
\end{equation}
In the provided context, $f(Q)$ represents an arbitrary function of $Q$, which is the non-metricity scalar. $\mathcal{L}_m$ stands for the matter Lagrangian density, $\mathcal{L}$ represents the Lagrangian density of the monopole, and $g$ denotes the determinant of the metric tensor $g_{\mu\nu}$.\\
The Lagrangian density for a self-coupled scalar triplet $\phi^a$ is given by:
\begin{equation}\label{1a}
{\mathcal {L}}= - \frac{\lambda }{4}(\phi ^2 -\eta ^2)^2 -\frac{1}{2} \sum _a g^{ij} \partial _i \phi ^a \partial _j \phi ^a.
\end{equation}
In this expression, $a = 1, 2, 3$, with $\eta$ and $\lambda$ representing the scales of gauge-symmetry breaking and self-interaction terms, respectively. The monopole's field configuration is
\begin{equation}\label{1b}
\phi ^a = \frac{\eta }{r} F(r) x^a\,.
\end{equation}
Here, the variable $x^a$ is defined as $(r \sin\theta \cos\phi, r \sin\theta \sin\phi, r \cos\theta)$, ensuring that $\sum _a x^a x^2 = r^2$. The Lagrangian density can then be expressed in terms of $F(r)$ by applying the field configuration as follows:
\begin{equation}\label{1c}
\mathcal{L} = -\left( 1-\frac{b(r)}{r}\right) \frac{\eta ^2 (F')^2}{2} -\frac{\eta ^2 F^2}{r^2}-\frac{\lambda \eta ^4}{4}(F^2-1)^2,
\end{equation}
and for the field $F(r)$, the Euler-Lagrange equation is given by
\begin{multline}\label{1d}
\left( 1-\frac{b(r)}{r}\right) F'' + F' \left[ \left( 1-\frac{b(r)}{r}\right) \frac{2}{r}+\frac{1}{2}\left( \frac{b-b' r}{r^2}\right) \right] \\
-F\left[ \frac{2}{r^2}+\lambda \eta ^2(F^2-1)\right] =0 .
\end{multline}
Also, the energy-momentum tensor is derived from Eq. \eqref{1a} as follows:
\begin{multline}\label{1e}
\bar{T}_{ij}=\partial _i\phi ^a\partial _j\phi ^a -\frac{1}{2} g_{ij}g^{\mu \nu }\partial _\mu \phi ^a\partial _\nu \phi ^a -\frac{g_{ij}\lambda }{4}(\phi ^2 -\eta ^2)^2.
\end{multline}
Thus, all four components of the Energy-Momentum tensor can be determined using Eq. \eqref{1e}:
\begin{equation}\label{1f}
\bar{T}^t_ t=-\eta ^2\left[ \frac{F^2}{r^2}+\left( 1-\frac{b(r)}{r}\right) \frac{(F')^2}{2} +\frac{\lambda \eta ^2}{4}(F^2-1)^2\right],
\end{equation}
\begin{equation}\label{1g}
\bar{T}^r_ r=-\eta ^2\left[ \frac{F^2}{r^2}+\left( 1-\frac{b(r)}{r}\right) \frac{(F')^2}{2} +\frac{\lambda \eta ^2}{4}(F^2-1)^2\right],
\end{equation}
\begin{multline}\label{1h}
\bar{T}^\theta _ \theta=\bar{T}^\Phi _\Phi \nonumber =-\eta ^2\left[ \left( 1-\frac{b(r)}{r}\right) \frac{(F')^2}{2} +\frac{\lambda \eta ^2}{4}(F^2-1)^2\right].
\end{multline}
It is difficult to get a precise analytical solution, as shown in Eq. \eqref{1d}. Hence, approximating the area outside the wormhole is sufficient to simplify and obtain the result. Consequently, the components of reduced energy-momentum are expressed as
\begin{equation}\label{1i}
\bar{T}^t_t=\bar{T}^r_r=-\frac{\eta ^2}{r^2} , \bar{T}^\theta _\theta =\bar{T}^\Phi _\Phi = 0.
\end{equation}
Additionally, the non-metricity tensor is defined by the equation given in \cite{J. B. Jimenez}
\begin{equation}\label{2}
Q_{\lambda\mu\nu}=\bigtriangledown_{\lambda} g_{\mu\nu}\,.
\end{equation}
Moreover, the superpotential, or non-metricity conjugate, is formally defined as follows:
\begin{equation}\label{3}
\hspace{-0.2cm}P^\alpha\,_{\mu\nu}=\frac{1}{4}\left[-Q^\alpha\;_{\mu\nu}+2Q_{(\mu}\;^\alpha\;_{\nu)}+Q^\alpha g_{\mu\nu}-\tilde{Q}^\alpha g_{\mu\nu}-\delta^\alpha_{(\mu}Q_{\nu)}\right].
\end{equation}
Traces of the non-metricity tensor can be expressed as follows:
\begin{equation}
\label{4}
\tilde{Q}_\alpha=Q^\mu\;_{\alpha\mu}\,,\;Q_{\alpha}=Q_{\alpha}\;^{\mu}\;_{\mu}.
\end{equation}
The non-metricity scalar is specified in \cite{J. B. Jimenez} as follows:
\begin{eqnarray}
\label{5}
Q &=& -P^{\alpha\mu\nu}\,Q_{\alpha\mu\nu}\\
&=& g^{\mu\nu}\left(L^\beta_{\,\,\,\alpha\beta}\,L^\alpha_{\,\,\,\mu\nu}-L^\beta_{\,\,\,\alpha\mu}\,L^\alpha_{\,\,\,\nu\beta}\right),
\end{eqnarray}
The disformation tensor, represented by $L^\beta_{\,\,\,\mu\nu}$, is defined as
\begin{equation}\label{6}
L^\beta_{\,\,\,\mu\nu}=\frac{1}{2}Q^\beta_{\,\,\,\mu\nu}-Q_{(\mu\,\,\,\,\,\,\nu)}^{\,\,\,\,\,\,\beta}.
\end{equation}
The equations of motion for gravity are derived by varying the action with respect to the metric tensor $g_{\mu\nu}$. This results in the following equation:
\begin{multline}\label{7}
\frac{-2}{\sqrt{-g}}\bigtriangledown_\alpha\left(\sqrt{-g}\,f_Q\,P^\alpha\;_{\mu\nu}\right)-\frac{1}{2}g_{\mu\nu}f \\
-f_Q\left(P_{\mu\alpha\beta}\,Q_\nu\;^{\alpha\beta}-2\,Q^
{\alpha\beta}\,\,_{\mu}\,P_{\alpha\beta\nu}\right)=8\pi T_{\mu\nu}\,,
\end{multline}
where $f_Q=\frac{\partial f}{\partial Q}$ and $T_{\mu\nu}$ represents the total energy-momentum tensor, which includes contributions from both the anisotropic fluid and the matter field. Thus, it can be expressed as:
\begin{equation}\label{8}
T_{\mu \nu } =\mathcal{T}_{\mu \nu } + \bar{T}_{\mu \nu }\,.
\end{equation}
The components of the energy-momentum tensor for the anisotropic fluid are given by:
\begin{equation}\label{9}
\mathcal{T}^\mu\;_\nu =\text{diag}(\rho,-p_r,-p_t,-p_t),
\end{equation}
Here, $\rho$ denotes the energy density, $p_r$ represents the radial pressure, and $p_t$ denotes the tangential pressure. The non-metricity scalar $Q$, related to the metric given in \eqref{11}, is defined according to the reference \cite{Tayde 4} as follows:
\begin{equation}\label{10}
Q=-\frac{b}{r^2}\left[2\phi^{'}+\frac{rb^{'}-b}{r(r-b)}\right].
\end{equation}\\
Thus, the field equations describing $f(Q)$ gravity for the wormhole, incorporating the Global Monopole Charge, can be written as:
\begin{multline}\label{12}
8 \pi  \rho =\frac{(r-b)}{2 r^3} \left[f_Q \left(\frac{(2 r-b) \left(r b'-b\right)}{(r-b)^2}+\frac{b \left(2 r \phi '+2\right)}{r-b}\right)
\right. \\ \left.
+\frac{f r^3}{r-b}+\frac{2 b r f_{\text{QQ}} Q'}{r-b}\right]-\frac{8 \pi  \eta ^2}{r^2},
\end{multline}
\begin{multline}\label{13}
8 \pi  p_r=-\frac{(r-b)}{2 r^3} \left[-f_Q \left(\frac{b }{r-b}\left(\frac{r b'-b}{r-b}+2+2 r \phi '\right)
\right.\right. \\ \left.\left.
-4 r \phi '\right)-\frac{f r^3}{r-b}-\frac{2 b r f_{\text{QQ}} Q'}{r-b}\right]+\frac{8 \pi  \eta ^2}{r^2},
\end{multline}
\begin{multline}\label{14}
8 \pi  p_t=\frac{(r-b)}{4 r^2} \left[f_Q \left(\frac{\left(r b'-b\right) \left(\frac{2 r}{r-b}+2 r \phi '\right)}{r (r-b)}+
\right.\right. \\ \left.\left.
\frac{4 (2 b-r) \phi '}{r-b}-4 r \left(\phi '\right)^2-4 r \phi ''\right)+\frac{2 f r^2}{r-b}
\right.\\\left.
-4 r f_{\text{QQ}} Q' \phi '\right].
\end{multline}
By employing these specific field equations, it becomes possible to thoroughly investigate various wormhole solutions within the context of $f(Q)$ gravity models.

\subsection{Energy conditions}
The classical energy conditions, which are derived from the Raychaudhuri equations, provide a foundational framework for discussing physically plausible matter configurations. Among the four commonly recognized energy conditions- namely weak, null, strong, and dominant-  the null energy condition holds particular importance, especially within the realm of GR when considering wormhole solutions. This significance arises from its direct correlation to the energy density required to sustain the openness of the wormhole throat. Any deviation from the null energy condition in the vicinity of the wormhole's throat suggests the presence of exotic matter characterized by negative energy density, a feature not typically associated with conventional matter sources. Energy conditions serve to constrain the stress-energy tensor, which describes the distribution of matter and energy in space-time and can be given as follows:\\
$\bullet$ The weak energy condition (\textbf{WEC}) :
$\rho\geq0$,\,\, $\rho+p_t\geq0$,\,\, and \,\, $\rho+p_r\geq0$.\\
$\bullet$ The null energy condition (\textbf{NEC}) : $\rho+p_t\geq0$\,\, and \,\, $\rho+p_r\geq0$.\\
$\bullet$ The dominant energy condition (\textbf{\textbf{DEC}}) :  $\rho\geq0$,\,\, $\rho+p_t\geq0$,\,\, $\rho+p_r\geq0$,\,\, $\rho-p_t\geq0$,\,\, and \,\, $\rho-p_r\geq0$.\\
$\bullet$ The strong energy condition (\textbf{SEC}) :
 $\rho+p_t\geq0$,\,\, $\rho+p_r\geq0$,\,\, and \,\, $\rho+p_r+2p_t\geq0$.\\
In summary, energy conditions are essential for constraining the behavior of matter in the Universe and are pivotal for our study of wormholes.
\section{Linear $f(Q)$ model}\label{sec3}
Now, in the current work, we assume a linear functional form of $f(Q)$ gravity, which is expressed as:
\begin{equation}\label{15}
f(Q)=\alpha Q\,,
\end{equation}
where $\alpha\neq0$ is a model parameter. This model parameter can be easily adjusted to retrieve GR for $\alpha=1$. This 
model is used in \cite{G. Mustafa} to study the traversable wormhole inspired by non-commutative geometries with conformal symmetry and found that wormhole solutions exist under this non-commutative geometry with viable physical properties. Hence, the reduced field equations with arbitrary redshift function $\phi(r)$ are given by
\begin{equation}\label{16}
\rho=\frac{\alpha  b'-8 \pi  \eta ^2}{8 \pi  r^2},
\end{equation}
\begin{equation}\label{17}
p_r = \frac{2 \alpha  r (b-r) \phi '+\alpha  b+8 \pi  \eta ^2 r}{8 \pi  r^3},
\end{equation}
\begin{equation}\label{18}
p_t = \frac{\alpha  \left(r \phi '+1\right) \left(r b'+2 r (b-r) \phi '-b\right)+2 \alpha  r^2 (b-r) \phi ''}{16 \pi  r^3}.
\end{equation}
Using the above field equations, we obtain the shape function $b(r)$ and study the energy conditions with the help of different DM profiles. These DM profiles will be elaborated in the next Section \ref{sec4}.
\section{Different Dark matter profiles}\label{sec4}
Within this section, we delve into the examination of two distinct profiles of DM, namely the Pseudo Isothermal (PI) and the Navarro-Frenk-White (NFW) profiles.
\subsection{Pseudo isothermal (PI) profile}\label{subsec1}
\begin{figure*}[t]
    \centering
    \includegraphics[width=17.5cm,height=6cm]{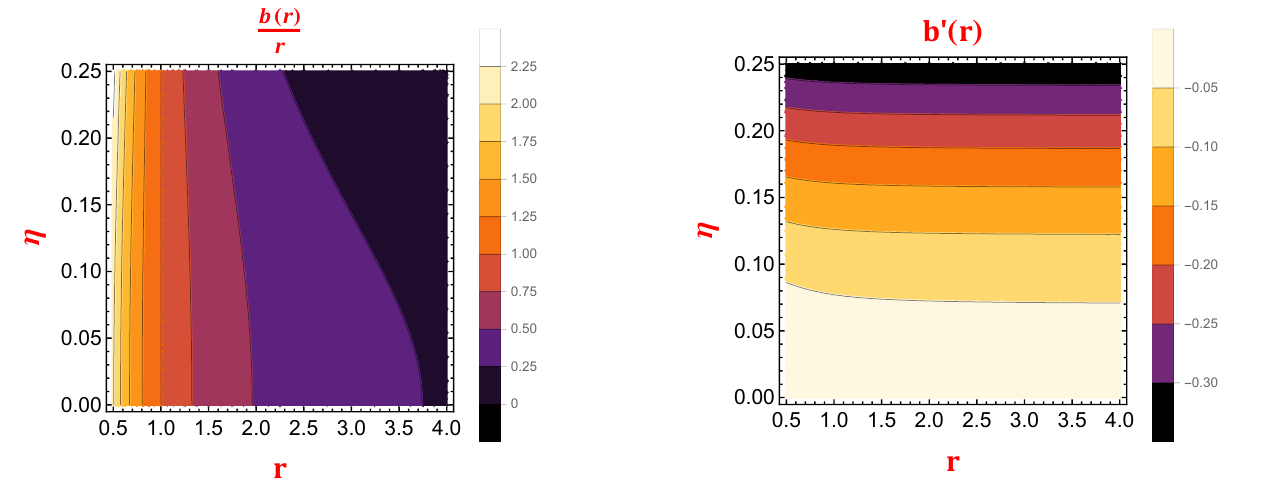}
    \caption{The contour plot displays PI profile with the variations in the asymptotic flatness condition \textit{(on the left)} and the flare-out condition \textit{(on the right)} as a function of the radial coordinate `$r$' under the redshift $\phi(r)=\text{constant}$. Furthermore, we keep other parameters fixed at constant values, including $\alpha=-5,\, r_s=0.5,\, \rho_s=0.02,\, \text{and} \, r_0 = 1$.}
    \label{fig1}
\end{figure*}

\begin{figure*}[t]
\centering
\includegraphics[width=14.5cm,height=6cm]{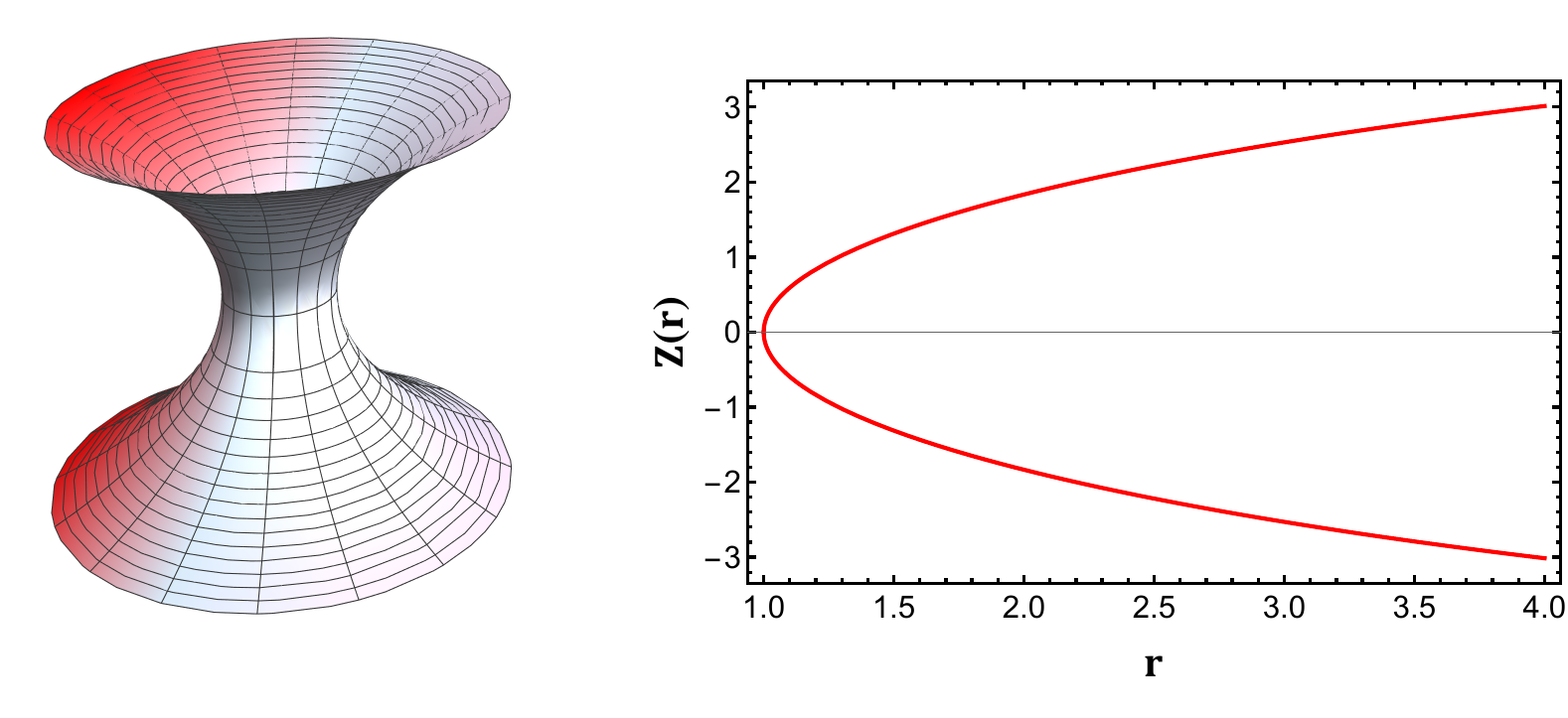}
\caption{The figure displays the embedding diagram for the PI profile. Furthermore, we keep other parameters fixed at constant values, including $\alpha=-5,\, r_s=0.5,\, \rho_s=0.02,\, \eta=0.15,\,\text{and} \, r_0 = 1$.}
\label{fig9}
\end{figure*}

One of the important classes of DM models is associated with modified gravity, such as Modified Newtonian Dynamics (MOND). In the MOND model, the DM density profile is described by the pseudo-isothermal profile \cite{K. G. Begeman}
\begin{equation}\label{19}
\rho=\frac{\rho_s}{\left(\frac{r}{r_s}\right)^2+1}\,,
\end{equation}
where $\rho_s$ is the central DM density and $r_s$ is the scale radius. In \cite{Zhaoyi Xu}, the author explored the possibility of traversable wormhole formation in the DM halos and obtained the exact solutions of the spherical symmetry traversable wormhole with isotropic pressure condition and found that WEC and NEC are violated at the throat of the wormhole. Moreover, author \cite{B. C. Paul} explored the existence of traversable wormholes in Einstein's general theory of relativity with density profile obtained from MOND with or without a scalar field.\\
Now, by comparing the energy densities, i.e. Eqs. \eqref{16} and \eqref{19}, we get the explicit form of shape function using the throat condition $b(r_0)=r_0$ as
\begin{multline}\label{20}
b(r)=\frac{1}{\alpha }\left(8 \pi  \rho_s {r_s}^3 \left(\tan ^{-1}\left(\frac{r_0}{{r_s}}\right)-\tan ^{-1}\left(\frac{r}{{r_s}}\right)\right)
\right. \\ \left.
+r_0 \left(\alpha -8 \pi  \left(\eta ^2+\rho_s {r_s}^2\right)\right)+8 \pi  r \left(\eta ^2+\rho_s {r_s}^2\right)\right)\,.
\end{multline}
Now, we will explore the visual representation of the shape function and examine the crucial conditions for the existence of a wormhole. For this purpose, we will meticulously choose the appropriate parameters. We showcase the graphical depiction of the asymptotic behavior of the shape function and the satisfaction of the flaring out condition in contour plot \ref{fig1} for the parameter $\eta$. The left contour in Fig. \ref{fig1} offers insights into the asymptotic nature of the shape function concerning the parameter $\eta$. As the radial distance increases, the ratio $\frac{b(r)}{r}$ tends toward $0$, confirming the asymptotic flatness behavior of the shape function. Moreover, the corresponding right graph vividly demonstrates the fulfillment of the flaring out condition, where $b'(r_0) < 1$, at the wormhole throat. Here, we consider the wormhole throat at $r_0=1$.\\
Further, following the formulation proposed by Morris and Thorne \cite{M. S. Morris}, the wormhole's embedding surface is described by the function $z(r)$, which is governed by the following differential equation:
\begin{equation}\label{6d1}
\frac{dz}{dr}=\pm \frac{1}{\sqrt{\frac{r}{b(r)}-1}}.
\end{equation}
The equation highlights that $\frac{dz}{dr}$ experiences divergence at the wormhole throat, indicating that the embedding surface assumes a vertical orientation at this crucial point. Moreover, Eq. \eqref{6d1} establishes the following relationship:
\begin{equation}\label{6d2}
z(r)=\pm \int_{r_0}^{r} \frac{dr}{\sqrt{\frac{r}{b(r)}-1}}.
\end{equation}
It is noted that the above integral cannot be solved analytically. Therefore, we resort to numerical techniques to generate the shape of the wormhole. Additionally, we present the embedding diagram $z(r)$ using Eq. \eqref{6d2}, as illustrated in Fig. \ref{fig9}.\\
In this investigation, we explore three distinct forms of the redshift function to study the energy conditions.

\subsubsection{$\phi(r)=c$}\label{subsubsec1}
The function $\phi(r) = c$ \cite{N. Godani, Zinnat 1}, where $c$ is a constant, is considered a suitable choice as a redshift function due to its consistent behavior concerning the radial coordinate $r$. It is worth noting that when $c = 0$, this function is termed the tidal force. For our analysis, we will proceed with positive values of $c$. Employing this redshift function alongside the shape function provided in Eq. \eqref{20} allows for the simultaneous derivation of explicit expressions for radial and tangential pressures, given by:
\begin{multline}\label{21}
p_r = \frac{1}{8 \pi  r^3}\left(8 \pi  \rho_s {r_s}^3 \left(\tan ^{-1}\left(\frac{r_0}{{r_s}}\right)-\tan ^{-1}\left(\frac{r}{{r_s}}\right)\right)
\right. \\ \left.
\hspace{0.7cm}+r_0 \left(\alpha -8 \pi  \left(\eta ^2+\rho_s {r_s}^2\right)\right)+8 \pi  r \left(2 \eta ^2+\rho_s {r_s}^2\right)\right)\,,
\end{multline}
\begin{multline}\label{22}
p_t = \frac{1}{16 r^3}\left(-\frac{r_0 \alpha }{\pi }+8 \left(\tan ^{-1}\left(\frac{r}{{r_s}}\right)-\tan ^{-1}\left(\frac{r_0}{{r_s}}\right)\right)
\right. \\ \left.
\hspace{0.7cm}\times \rho_s {r_s}^3 +8 r_0 \left(\eta ^2+\rho_s {r_s}^2\right)-\frac{8 \rho_s r {r_s}^4}{r^2+{r_s}^2}\right)\,.
\end{multline}
Additionally, NEC at the throat of wormhole $r=r_0$ is given by
\begin{equation}
\left(\rho + p_r\right)_{\text{at}\,\, r=r_0}=\frac{\frac{\alpha }{8 \pi }+\eta ^2}{r_0^2}+\frac{\rho_s r_s^2}{r_0^2+r_s^2}\,,
\end{equation}
\begin{equation}
\left(\rho + p_t\right)_{\text{at}\,\, r=r_0}=\frac{3 \rho_s r_s^2}{2 \left(r_0^2+r_s^2\right)}-\frac{\alpha -8 \pi  \eta ^2}{16 \pi  r_0^2}\,.
\end{equation}
Additionally, we present the graphical illustration of this condition in Fig. \ref{fig2}, providing a visual understanding of the radial and tangential pressures at the wormhole throat.
\subsubsection{$\phi(r)=\frac{1}{r}$}\label{subsubsec2}
The function $\phi(r) = \frac{1}{r}$ \cite{S. Kar, F. Parsaei}, is another appropriate option as a redshift function due to its regular behavior for $r > 0$, effectively avoiding the event horizon beyond the wormhole throat. Utilizing this redshift function, the expressions for the radial and tangential pressures can be formulated as follows:
\begin{multline}\label{23}
p_r = \frac{1}{8 \pi  r^4}\left(r_0 (r-2) \left(\alpha -8 \pi  \left(\eta ^2+\rho_s {r_s}^2\right)\right)+8 \pi \rho_s {r_s}^3
\right. \\ \left.
\hspace{0.7cm}\times (r-2)  \left(\tan ^{-1}\left(\frac{r_0}{{r_s}}\right)-\tan ^{-1}\left(\frac{r}{{r_s}}\right)\right)+2 r \left(\alpha +8 \pi  
\right.\right. \\ \left.\left.
\times \eta ^2 (r-1)+4 \pi  \rho_s (r-2) {r_s}^2\right)\right)\,,
\end{multline}
\begin{multline}\label{24}
p_t = \frac{1}{16 r^5}\left(\frac{r_0}{\pi } \left((r-3) r-2\right) \left(8 \pi  \left(\eta ^2+\rho_s {r_s}^2\right)-\alpha \right)
\right. \\ \left.
\hspace{0.7cm}+8 \rho_s ((r-3) r-2) {r_s}^3 \left(\tan ^{-1}\left(\frac{r}{{r_s}}\right)-\tan ^{-1}\left(\frac{r_0}{{r_s}}\right)\right)
\right. \\ \left.
\hspace{0.8cm}+\frac{8 \rho_s r {r_s}^2 }{r^2+{r_s}^2}\left(2 (r+1) r^2+(2-(r-3) r) {r_s}^2\right)-\frac{2 r }{\pi }
\right. \\ \left.
\times (r+1) \left(\alpha -8 \pi  \eta ^2\right)\right)\,.
\end{multline}
Also, NEC for radial and tangential pressures at the throat of the wormhole is given by
\begin{equation}
\left(\rho + p_r\right)_{\text{at}\,\, r=r_0}=\frac{\frac{\alpha }{8 \pi }+\eta ^2}{r_0^2}+\frac{\rho_s r_s^2}{r_0^2+r_s^2}\,,
\end{equation}
\begin{multline}
\left(\rho + p_t\right)_{\text{at}\,\, r=r_0}=\frac{1}{16r_0^3}\left(\frac{8 r_0^2 (3 r_0-1) \rho_s r_s^2}{r_0^2+r_s^2}
\right. \\ \left.
+\frac{(r_0-1) \left(8 \pi  \eta ^2-\alpha \right)}{\pi }\right)\,,
\end{multline}
Moreover, the concept can be effectively illustrated through a graphical representation, as shown in Fig. \ref{fig3}.
\subsubsection{$\phi(r)=\log\left(1+\frac{r_0}{r}\right)$}\label{subsubsec3}
The function $\phi(r)=\log\left(1+\frac{r_0}{r}\right)$ \cite{N. Godani}, which is also suitable as a redshift function, and it avoids the event horizon after the wormhole throat. For this redshift function, the radial and tangential pressures can be given as
\begin{multline}\label{25}
p_r = \frac{1}{8 \pi  r^3 (r_0+r)}\left(-r_0^2 \alpha +8 \pi  r_0^2 \eta ^2+3 r_0 \alpha  r-8 \pi  r_0 \eta ^2 r
\right. \\ \left.
\hspace{0.7cm}+8 \pi  \rho_s {r_s}^2 (r_0-r)^2+8 \pi  \rho_s {r_s}^3 (r_0-r) \left(\tan ^{-1}\left(\frac{r}{{r_s}}\right)
\right.\right. \\ \left.\left.
\hspace{0.7cm}-\tan ^{-1}\left(\frac{r_0}{{r_s}}\right)\right)+16 \pi  \eta ^2 r^2\right)\,,
\end{multline}
\begin{multline}\label{26}
p_t = \frac{1}{16 \pi  r^3 (r_0+r) \left(r^2+{r_s}^2\right)}\left(-8 \pi  \rho_s {r_s}^2 \left(2 r_0^2 \left(r^2+r_s^2\right)
\right.\right. \\ \left.\left.
\hspace{0.7cm}-3 r_0 r \left(r^2+{r_s}^2\right)+r^2 {r_s}^2\right)+r_0 \left(\alpha -8 \pi  \eta ^2\right) (2 r_0-3 r) 
\right. \\ \left.
\hspace{0.7cm}\times\left(r^2+{r_s}^2\right)+ \left(\tan ^{-1}\left(\frac{r_0}{{r_s}}\right)-\tan ^{-1}\left(\frac{r}{{r_s}}\right)\right)8 \pi  \rho_s
\right. \\ \left.
\hspace{0.8cm}\times{r_s}^3 (2 r_0-r) \left(r^2+{r_s}^2\right)\right)\,.
\end{multline}
Moreover, NEC at the wormhole throat can be given by
\begin{equation}
\left(\rho + p_r\right)_{\text{at}\,\, r=r_0}=\frac{\frac{\alpha }{8 \pi }+\eta ^2}{r_0^2}+\frac{\rho_s r_s^2}{r_0^2+r_s^2}\,,
\end{equation}
\begin{equation}
\left(\rho + p_t\right)_{\text{at}\,\, r=r_0}=\frac{5 \rho_s r_s^2}{4 \left(r_0^2+r_s^2\right)}-\frac{\alpha -8 \pi  \eta ^2}{32 \pi  r_0^2}\,.
\end{equation}
Furthermore, the concept can be aptly depicted through a graphical representation, as demonstrated in Fig. \ref{fig4}.
\begin{figure*}[t]
\centering
\includegraphics[width=17.5cm,height=5cm]{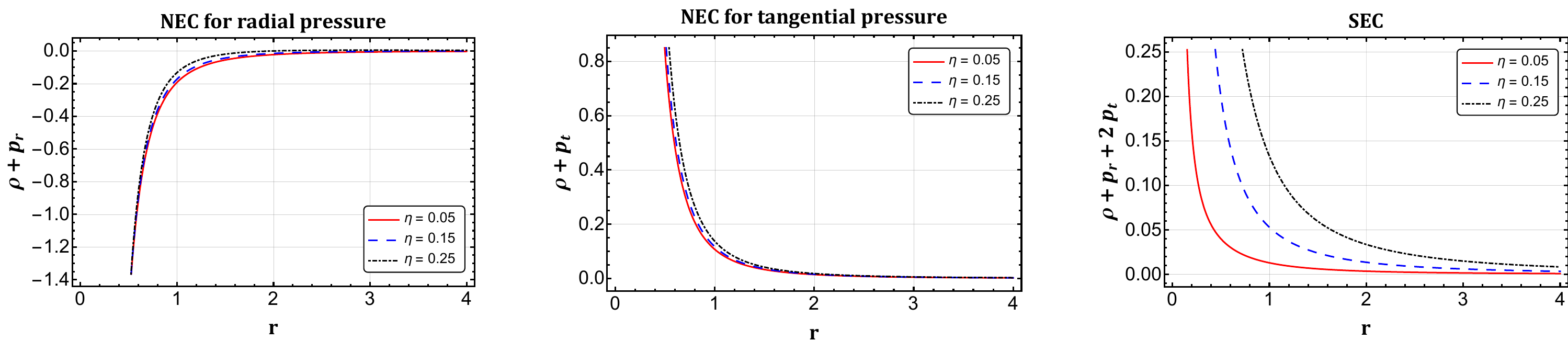}
\caption{The figure displays the PI profile with the variations in the NEC for both pressure and SEC as a function of the radial coordinate `$r$' for various values of `$\eta$ ' under the redshift $\phi(r)=\text{constant}$. Furthermore, we keep other parameters fixed at constant values, including $\alpha=-5,\, r_s=0.5,\, \rho_s=0.02,\, \text{and} \, r_0 = 1$.}
    \label{fig2}
\end{figure*}
\begin{figure*}[t]
    \centering
    \includegraphics[width=17.5cm,height=5cm]{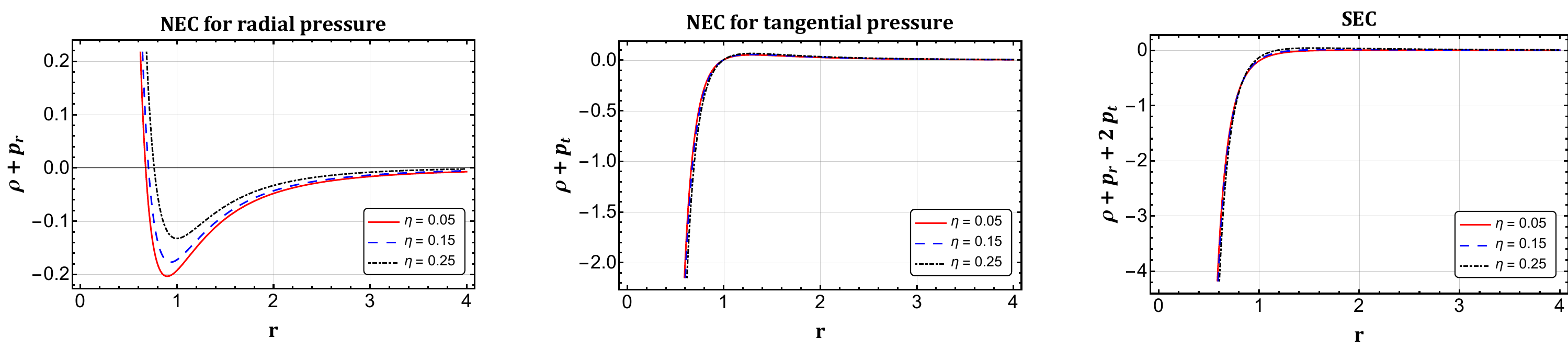}
    \caption{The figure displays the PI profile with the variations in the NEC for both pressure and SEC as a function of the radial coordinate `$r$' for various values of `$\eta$ ' under the redshift $\phi(r)=\frac{1}{r}$. Furthermore, we keep other parameters fixed at constant values, including $\alpha=-5,\, r_s=0.5,\, \rho_s=0.02,\, \text{and} \, r_0 = 1$.}
    \label{fig3}
\end{figure*}
\begin{figure*}[t]
    \centering
    \includegraphics[width=17.5cm,height=5cm]{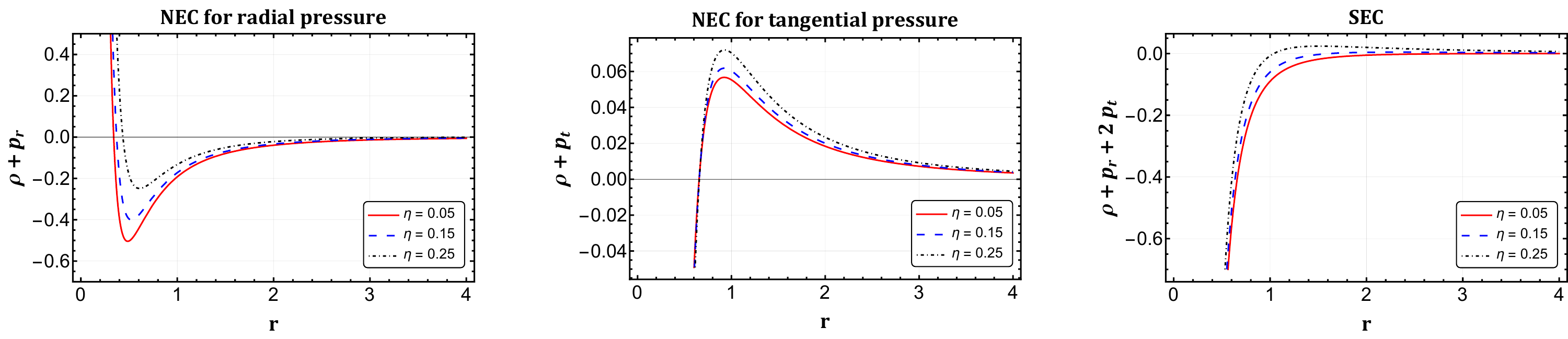}
    \caption{The figure displays the PI profile with the variations in the NEC for both pressure and SEC as a function of the radial coordinate `$r$' for various values of `$\eta$ ' under the redshift $\phi(r)=\log\left(1+\frac{r_0}{r}\right)$. Furthermore, we keep other parameters fixed at constant values, including $\alpha=-5,\, r_s=0.5,\, \rho_s=0.02,\, \text{and} \, r_0 = 1$.}
    \label{fig4}
\end{figure*}
\begin{table*}[t]
\begin{tabular}{cllll}
\hline\hline
 & \multicolumn{3}{p{12.5cm}}{Psuedo Isothermal profile ($\alpha=-5,\, r_s=0.5,\, \rho_s=0.02,\, \eta=0.05,\,\text{and} \, r_0 = 1$)} \\ \hline\hline
 &   \multicolumn{1}{l}{\hspace{0.5cm}$\phi(r)=c$}  &  \multicolumn{1}{l}{\hspace{0.5cm}$\phi(r)=\frac{1}{r}$}   &  \multicolumn{1}{l}{\hspace{0.5cm}$\phi(r)=\log\left(1+\frac{r_0}{r}\right)$} \\ \hline
$\rho$  & $>0$ for all $r$   &  $>0$ for all $r$      &  $>0$ for all $r$  \\\\ 
$\rho + p_r$ &  \multicolumn{1}{p{3cm}}{$<0$ for $r\in(0,13.9)$ $>0$ for $r\in[13.9,\infty)$}&     \multicolumn{1}{p{4cm}}{$>0$ for $r\in(0,0.7)$$\cup [40.7,\infty)$ $<0$ for $r\in[0.7,40.7)$}      &  \multicolumn{1}{p{4cm}}{$>0$ for $r\in(0,0.3]$$\cup (40,\infty)$ $<0$ for $r\in(0.3,40]$}     \\\\ 
$\rho + p_t$ &  \multicolumn{1}{p{3cm}}{$>0$ for $r\in(0,\infty)$ }   &    \multicolumn{1}{p{2.5cm}}{$<0$ for $r\in(0,1)$ \hspace{0.5cm} $>0$ for $r\in[1,\infty)$}       &  \multicolumn{1}{p{2.7cm}}{$<0$ for $r\in(0,0.6]$ \hspace{0.5cm} $>0$ for $r\in(0.6,\infty)$}   \\ \\
$\rho - p_r$ &   \multicolumn{1}{p{3cm}}{$>0$ for $r\in(0,41.6)$  $<0$ for $r\in[41.6,\infty)$}  &   \multicolumn{1}{p{4cm}}{$<0$ for $r\in(0,0.7)$$\cup [124,\infty)$ $>0$ for $r\in[0.7,124)$}        &  \multicolumn{1}{p{4.3cm}}{$<0$ for $r\in(0,0.34)$$\cup (122,\infty)$ $>0$ for $r\in[0.34,122]$}  \\ \\
$\rho - p_t$ &  \multicolumn{1}{p{3cm}}{$<0$ for $r\in(0,20.8)$  $>0$ for $r\in[20.8,\infty)$}   &    \multicolumn{1}{p{4cm}}{$>0$ for $r\in(0,1]$$\cup (61.6,\infty)$ $<0$ for $r\in(1,61.6]$}       &   \multicolumn{1}{p{4.3cm}}{$>0$ for $r\in(0,0.67]$$\cup (60.3,\infty)$ $<0$ for $r\in(0.67,60.3]$}    \\ \\
$\rho + p_r + 2p_t$ &  \multicolumn{1}{p{3cm}}{$>0$ for $r\in(0,\infty)$ }   &   \multicolumn{1}{p{2.7cm}}{$<0$ for $r\in(0,1.6]$ \hspace{0.5cm} $>0$ for $r\in(1.6,\infty)$}         &    \multicolumn{1}{p{2.7cm}}{$<0$ for $r\in(0,3.2]$ \hspace{0.5cm} $>0$ for $r\in(3.2,\infty)$}   \\ \hline\hline
\end{tabular}
\caption{Summary for results of PI profile for three different redshift functions}
\label{table:1}
\end{table*}
\subsection{NFW profile}\label{subsec2}
\begin{figure*}[t]
    \centering
    \includegraphics[width=17.5cm,height=6cm]{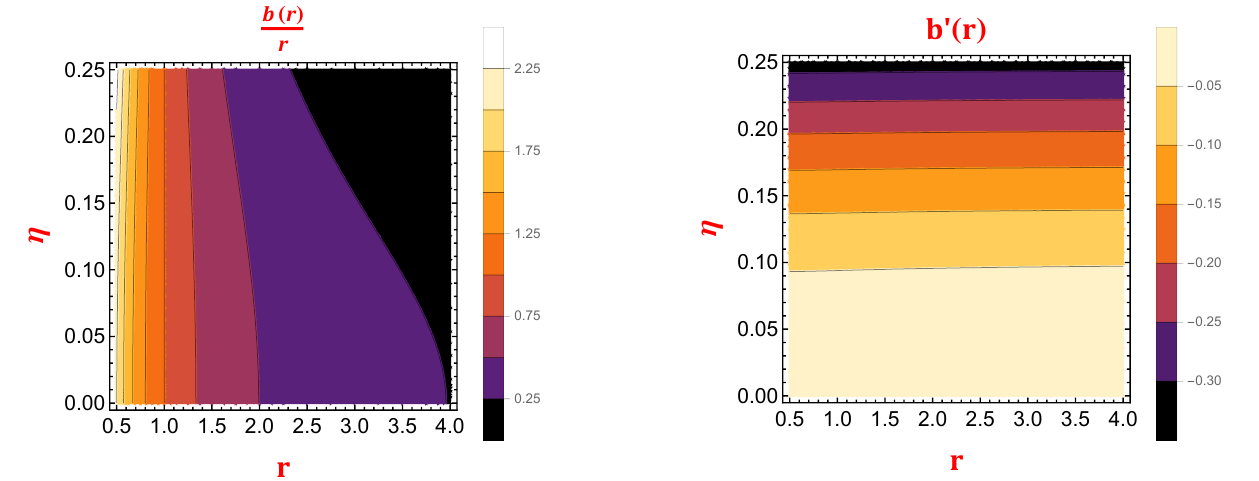}
    \caption{The contour plot displays NFW profile with the variations in the asymptotic flatness condition \textit{(on the left)} and the flare-out condition \textit{(on the right)} as a function of the radial coordinate `$r$' under the redshift $\phi(r)=\text{constant}$. Furthermore, we keep other parameters fixed at constant values, including $\alpha=-5,\, r_s=0.5,\, \rho_s=0.02,\, \text{and} \, r_0 = 1$.}
    \label{fig5}
\end{figure*}

\begin{figure*}[t]
\centering
\includegraphics[width=14.5cm,height=6cm]{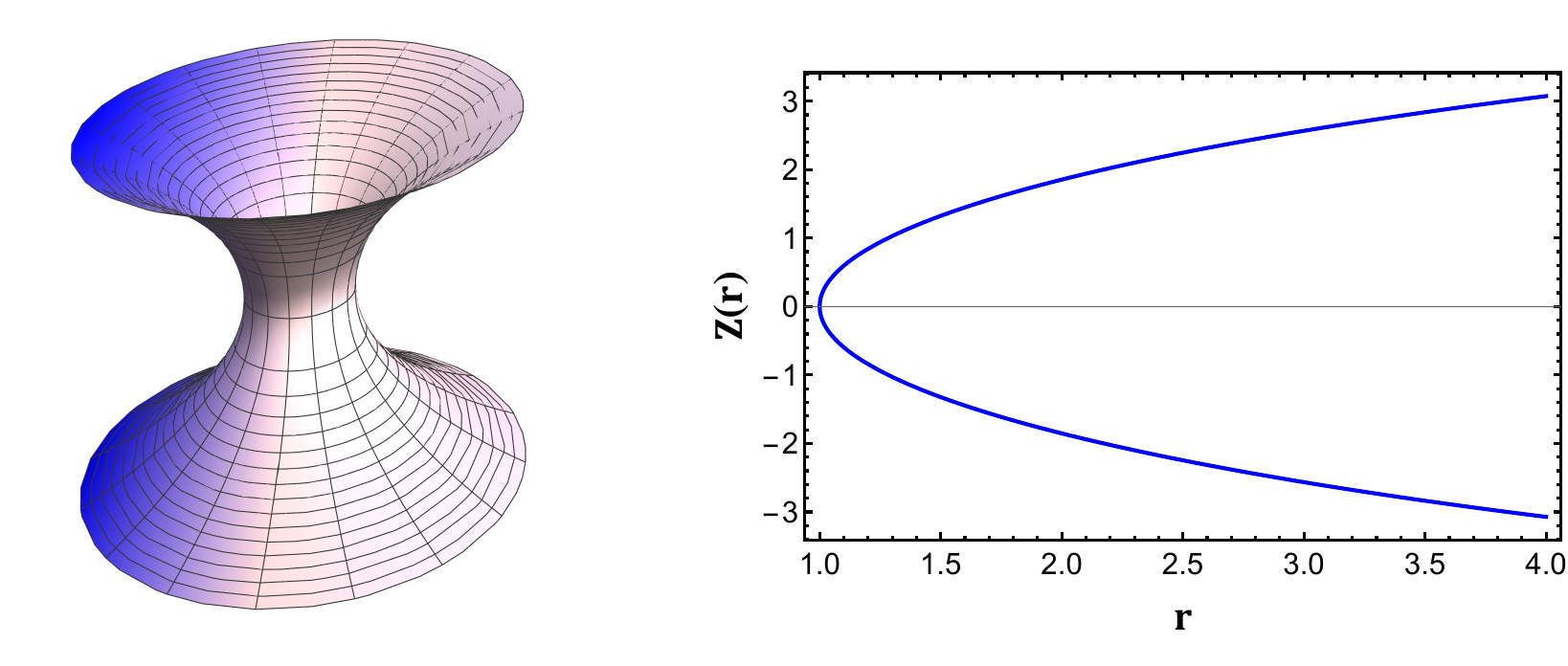}
\caption{The figure displays the embedding diagram for the NFW profile. Furthermore, we keep other parameters fixed at constant values, including $\alpha=-5,\, r_s=0.5,\, \rho_s=0.02,\, \eta=0.15,\,\text{and} \, r_0 = 1$.}
\label{fig10}
\end{figure*}

The potential density model introduced by Hernquist \cite{L. Hernquist} aimed to investigate both theoretical and observational aspects of elliptical galaxies. Subsequently, Navarro and his colleagues \cite{J. F. Navarro} analyzed equilibrium density profiles of DM halos in universes exhibiting hierarchical clustering through high-resolution N-body simulations. This model delineates how the density of DM varies concerning the distance from the center of a halo. Their studies demonstrated that the structure of these profiles is consistent regardless of the halo mass, the spectral shape of the initial density fluctuations, or the cosmological parameter values. The Cold DM halo models for X-ray clusters and elliptical galaxies, as defined by Navarro and colleagues \cite{J. F. Navarro}, are as follows:
\begin{equation}\label{27}
\rho = \frac{\rho_s r_s}{r \left(\frac{r}{{r_s}}+1\right)^2}\,,
\end{equation}
where $r_s$ and $\rho_s$ denote the characteristic radius and central density of the Universe, respectively.
Now, by comparing the energy densities given by Eqs. \eqref{16} and \eqref{27}, we get the explicit form of shape function using the throat condition $b(r_0)=r_0$ as
\begin{multline}\label{28}
b(r)=-\frac{8 \pi}{\alpha }  \left(\frac{\rho_s {r_s}^4}{{r_0}+{r_s}}+\rho_s {r_s}^3 \log ({r_0}+{r_s})+\eta ^2 ({r_0}+{r_s})\right)
\\
\hspace{0.5cm}+{r_0}+\frac{8 \pi  }{\alpha }\left(\frac{\rho_s {r_s}^4}{r+{r_s}}+\rho_s {r_s}^3 \log (r+{r_s})+\eta ^2 (r+{r_s})\right)\,.
\end{multline}
We will now elucidate the graphical representation of the derived shape function as well as the requirements required for a wormhole to exist. To achieve this, we carefully select appropriate parameters. Initially, we will investigate the behavior of the shape function conditions for the NFW profile. Contour plot \ref{fig5} visualizes the asymptotic flatness condition and the flaring out condition for the parameter $\eta$. The left contour in Fig. \ref{fig5} illustrates the asymptotic behavior of the shape function concerning the parameter $\eta$. It is observed that as the radial distance increases, the ratio $\frac{b(r)}{r}$ tends toward zero, verifying the asymptotic behavior of the shape function. The corresponding right graph demonstrates the satisfaction of the flaring out condition, denoted as ``$b'(r_0) < 1$" at the wormhole throat. Here, we consider the wormhole throat at $r_0=1$.\\
Moving on, for the NFW profile wormhole shape function \eqref{28}, we present the embedding diagram $z(r)$ and its comprehensive visualization in Fig. \ref{fig10}. Again, we explore three distinct forms of the redshift function for this profile to study the energy conditions.
\subsubsection{$\phi(r)=c$}\label{subsubsec4}
Using this particular form of the redshift function $\phi(r)=c$ and the shape function provided by \eqref{28}, the radial \eqref{17} and tangential \eqref{18} pressures are given by
\begin{multline}\label{29}
p_r =\frac{1}{8 r^3}\left(\frac{r_0 \alpha }{\pi }-8 \eta ^2 (r_0-2 r)+\frac{8 \rho_s {r_s}^4 (r_0-r)}{(r_0+{r_s}) (r+{r_s})}
\right. \\ \left.
+8 \rho_s {r_s}^3 (\log (r+{r_s})-\log (r_0+{r_s}))\right)\,,
\end{multline}
\begin{multline}\label{30}
p_t = \frac{1}{16 r^3}\left(\frac{8 \rho_s {r_s}^3 \left(r_0 \left(r^2-r {r_s}-{r_s}^2\right)+r {r_s} (2 r+{r_s})\right)}{(r_0+{r_s}) (r+{r_s})^2}
\right. \\ \left.
\hspace{1cm}+r_0 \left(8 \eta ^2-\frac{\alpha }{\pi }\right)+8 \left(\log (r_0+{r_s})-\log (r+{r_s})\right)
\right. \\ \left.
\times \rho_s {r_s}^3 \right)\,.
\end{multline}
Furthermore, NEC for radial and tangential pressures at the
wormhole throat is given by
\begin{equation}
\left(\rho + p_r\right)_{\text{at}\,\, r=r_0}=\frac{1}{8r_0^2}\left(8 \left(\frac{r_0 \rho_s r_s^3}{(r_0+r_s)^2}+\eta ^2\right)+\frac{\alpha }{\pi }\right)\,,
\end{equation}
\begin{equation}
\left(\rho + p_t\right)_{\text{at}\,\, r=r_0}=\frac{1}{16r_0^2}\left(8 \left(\frac{3r_0 \rho_s r_s^3}{(r_0+r_s)^2}+\eta ^2\right)-\frac{\alpha }{\pi }\right)\,.
\end{equation}
And its graphical representation is shown in Fig. \ref{fig6}.
\subsubsection{$\phi(r)=\frac{1}{r}$}\label{subsubsec5}
Utilizing this specific form of the redshift function $\phi(r)=\frac{1}{r}$ and the shape function provided by Eq. \eqref{28}, the expressions for the radial \eqref{17} and tangential \eqref{18} pressures are given by:
\begin{multline}\label{31}
p_r = \frac{1}{8 r^4}\left(\frac{r_0}{\pi } (r-2) \left(\alpha -8 \pi  \eta ^2\right)+\frac{2 r}{\pi } \left(\alpha +8 \pi  \eta ^2 (r
\right.\right. \\ \left.\left.
\hspace{0.8cm}-1)\right)+\frac{8 \rho_s (r-2) {r_s}^4 (r_0-r)}{(r_0+{r_s}) (r+{r_s})}+8 \rho_s (r-2) {r_s}^3 
\right. \\ \left.
\times (\log (r+{r_s})-\log (r_0+{r_s}))\right)\,,
\end{multline}
\begin{multline}\label{32}
p_t = \frac{1}{16 \pi  r^5 (r_0+{r_s}) (r+{r_s})^2}\left(-r_0^2 ((r-3) r-2) \left(\alpha 
\right.\right. \\ \left.\left.
\hspace{0.7cm} -8 \pi  \eta ^2\right)(r+{r_s})^2+r_0 \left(8 \pi  \rho_s {r_s}^3 \left((r-1) r^3+(2
\right.\right.\right. \\ \left.\left.\left.
\hspace{0.7cm}-(r-3) r) {r_s}^2+(2-(r-3) r) r {r_s}\right)-\left(\alpha -8 \pi  \eta ^2\right)  
\right.\right. \\ \left.\left.
\hspace{0.7cm}\times (r+{r_s})^2 (((r-3) r-2){r_s}+2 r (r+1))\right)+8 \pi  \rho_s 
\right. \\ \left.
\hspace{0.7cm}\times ((r-3) r-2) {r_s}^3 (r_0+{r_s}) (r+{r_s})^2 (\log (r_0+{r_s})
\right. \\ \left.
\hspace{0.7cm}-\log (r+{r_s}))+8 \pi  \rho_s r {r_s}^4 (r (r (2 r+{r_s}-4)-3 {r_s}
\right. \\ \left.
-2)-2 {r_s})-2 r (r+1) {r_s} \left(\alpha -8 \pi  \eta ^2\right) (r+{r_s})^2\right)\,.
\end{multline}
Moreover, it is essential to consider the NEC for the radial and tangential pressures at the wormhole throat, which can be expressed as:
\begin{equation}
\left(\rho + p_r\right)_{\text{at}\,\, r=r_0}=\frac{1}{8r_0^2}\left(8 \left(\frac{r_0 \rho_s r_s^3}{(r_0+r_s)^2}+\eta ^2\right)+\frac{\alpha }{\pi }\right)\,,
\end{equation}
\begin{multline}
\left(\rho + p_t\right)_{\text{at}\,\, r=r_0}=\frac{1}{16r_0^3}\left(\frac{(r_0-1) \left(8 \pi  \eta ^2-\alpha \right)}{\pi }
\right. \\ \left.
+\frac{8 r_0 (3 r_0-1) \rho_s r_s^3}{(r_0+r_s)^2}\right)\,.
\end{multline}
Additionally, we present the graphical illustration of this condition in Fig. \ref{fig7}, providing a visual understanding of the radial and tangential pressures at the wormhole throat.
\subsubsection{$\phi(r)=\log\left(1+\frac{r_0}{r}\right)$}\label{subsubsec6}
With the help of the specific form of the redshift function $\phi(r)=\log\left(1+\frac{r_0}{r}\right)$ alongside the shape function provided by Eq. \eqref{28}, we can derive the expressions for the radial \eqref{17} and tangential \eqref{18} pressures as follows:
\begin{multline}\label{33}
p_r = -\frac{1}{8 \pi  r^3 (r_0+r) (r_0+{r_s}) (r+{r_s})}\left((r_0+{r_s}) (r+{r_s}) 
\right. \\ \left.
\hspace{0.8cm}\times \left(r_0 \alpha (r_0-3 r)-8 \pi  \eta ^2 \left(r_0^2-r_0 r+2 r^2\right)\right)+8 \pi  \rho_s 
\right. \\ \left.
\hspace{0.8cm}\times {r_s}^4 (r_0-r)^2-8 \pi  \rho_s {r_s}^3(\log (r_0+{r_s})-\log (r+{r_s}))
\right. \\ \left.
 \times (r_0-r) (r_0+{r_s}) (r+{r_s}) \right)\,,
\end{multline}
\begin{multline}\label{34}
p_t = \frac{1}{16 \pi  r^3 (r_0+r) (r_0+{r_s}) (r+{r_s})^2}\left(8 \pi  \rho_s {r_s}^3 \left(2 r_0^2 {r_s} (r
\right.\right. \\ \left.\left.
\hspace{0.8cm}+{r_s})+r_0 r \left(r^2-3 r {r_s}-3 {r_s}^2\right)+r^2 {r_s} (2 r+{r_s})\right)-8 \pi 
\right. \\ \left.
\hspace{0.8cm}\times \rho_s {r_s}^3 (2 r_0-r) (r_0+{r_s}) (\log (r_0+{r_s})-\log (r+{r_s}))
\right. \\ \left.
 \times(r+{r_s})^2+r_0 \left(\alpha -8 \pi  \eta ^2\right) (2 r_0-3 r) (r_0+{r_s}) (r+{r_s})^2\right)\,.
\end{multline}
Also, it is crucial to examine the NEC for the radial and tangential pressures at the wormhole throat, which can be expressed as:
\begin{equation}
\left(\rho + p_r\right)_{\text{at}\,\, r=r_0}=\frac{1}{8r_0^2}\left(8 \left(\frac{a \text{$\rho $0} s^3}{(a+s)^2}+\eta ^2\right)+\frac{\alpha }{\pi }\right)\,,
\end{equation}
\begin{multline}
\left(\rho + p_t\right)_{\text{at}\,\, r=r_0}=\frac{1}{32r_0^2}\left(8 \left(\frac{5 a \text{$\rho $0} s^3}{(a+s)^2}+\eta ^2\right)-\frac{\alpha }{\pi }\right)\,.
\end{multline}
And it can be demonstrated through a graphical representation, as depicted in Fig. \ref{fig8}.
\begin{figure*}[t]
    \centering
    \includegraphics[width=17.5cm,height=5cm]{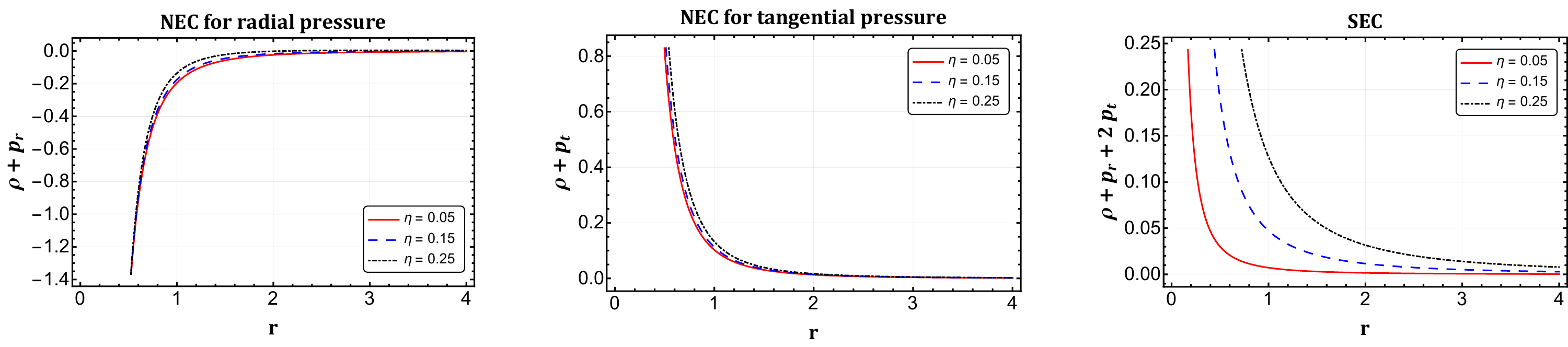}
    \caption{The figure displays the NFW profile with the variations in the NEC for both pressure and SEC as a function of the radial coordinate `$r$' for various values of `$\eta$ ' under the redshift $\phi(r)=\text{constant}$. Furthermore, we keep other parameters fixed at constant values, including $\alpha=-5,\, r_s=0.5,\, \rho_s=0.02,\, \text{and} \, r_0 = 1$.}
    \label{fig6}
\end{figure*}
\begin{figure*}[t]
    \centering
    \includegraphics[width=17.5cm,height=5cm]{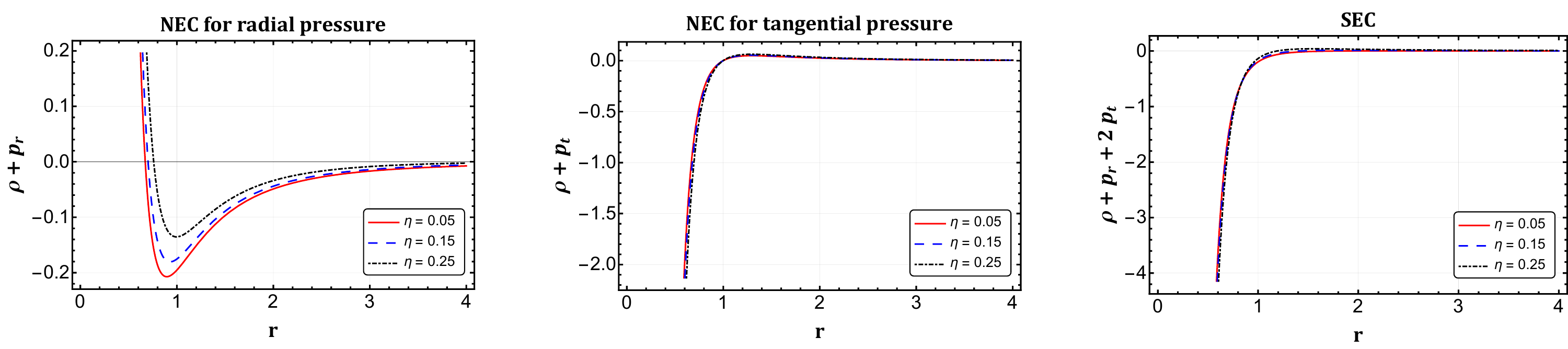}
    \caption{The figure displays the NFW profile with the variations in the NEC for both pressure and SEC as a function of the radial coordinate `$r$' for various values of `$\eta$ ' under the redshift $\phi(r)=\frac{1}{r}$. Furthermore, we keep other parameters fixed at constant values, including $\alpha=-5,\, r_s=0.5,\, \rho_s=0.02,\, \text{and} \, r_0 = 1$.}
    \label{fig7}
\end{figure*}
\begin{figure*}[t]
    \centering
    \includegraphics[width=17.5cm,height=5cm]{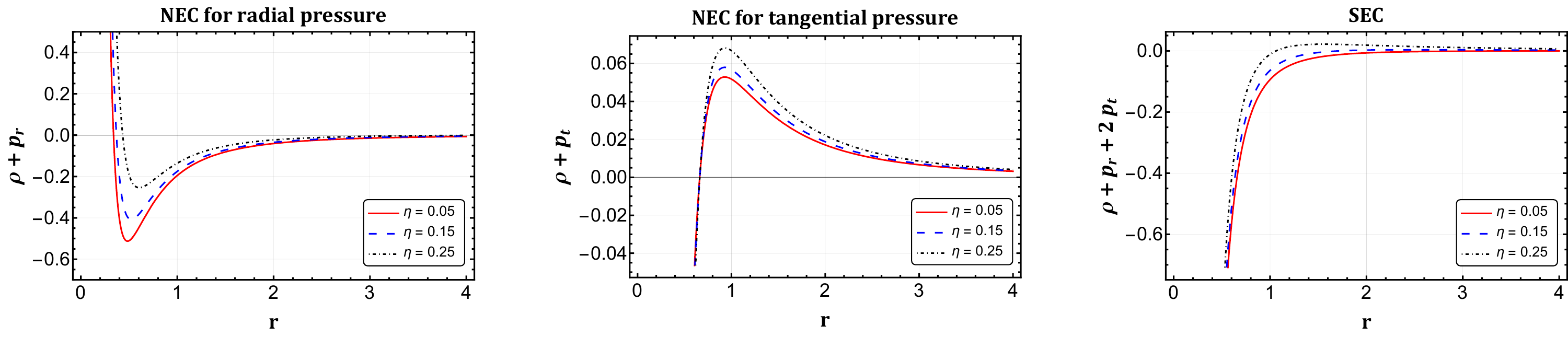}
    \caption{The figure displays the NFW profile with the variations in the NEC for both pressure and SEC as a function of the radial coordinate `$r$' for various values of `$\eta$ ' under the redshift $\phi(r)=\log\left(1+\frac{r_0}{r}\right)$. Furthermore, we keep other parameters fixed at constant values, including $\alpha=-5,\, r_s=0.5,\, \rho_s=0.02,\, \text{and} \, r_0 = 1$.}
    \label{fig8}
\end{figure*}
\begin{table*}[t]
\begin{tabular}{cllll}
\hline\hline
 & \multicolumn{3}{p{12.5cm}}{\hspace{1cm}NFW profile ($\alpha=-5,\, r_s=0.5,\, \rho_s=0.02,\, \eta=0.05,\,\text{and} \, r_0 = 1$)} \\ \hline\hline
 &   \multicolumn{1}{l}{\hspace{0.5cm}$\phi(r)=c$}  &  \multicolumn{1}{l}{\hspace{0.5cm}$\phi(r)=\frac{1}{r}$}   &  \multicolumn{1}{l}{\hspace{0.5cm}$\phi(r)=\log\left(1+\frac{r_0}{r}\right)$} \\ \hline
$\rho$  & $>0$ for all $r$   &  $>0$ for all $r$      &  $>0$ for all $r$  \\\\ 
$\rho + p_r$ &  \multicolumn{1}{p{3cm}}{$<0$ for $r\in(0,38.3]$ $>0$ for $r\in(38.3,\infty)$}&     \multicolumn{1}{p{4.5cm}}{$>0$ for $r\in(0,0.67]$$\cup (117.7,\infty)$ $<0$ for $r\in(0.67,117.7]$}      &  \multicolumn{1}{p{4.4cm}}{$>0$ for $r\in(0,0.33]$$\cup (117,\infty)$ $<0$ for $r\in(0.33,117]$}     \\\\ 
$\rho + p_t$ &  \multicolumn{1}{p{3cm}}{$>0$ for $r\in(0,\infty)$ }   &    \multicolumn{1}{p{3cm}}{$<0$ for $r\in(0,0.99]$ \hspace{0.5cm} $>0$ for $r\in(0.99,\infty)$}       &  \multicolumn{1}{p{2.9cm}}{$<0$ for $r\in(0,0.66]$ \hspace{0.5cm} $>0$ for $r\in(0.66,\infty)$}   \\ \\
$\rho - p_r$ &   \multicolumn{1}{p{3cm}}{$>0$ for $r\in(0,39.2]$  $<0$ for $r\in(39.2,\infty)$}  &   \multicolumn{1}{p{4.5cm}}{$<0$ for $r\in(0,0.67]$$\cup (118.6,\infty)$ $>0$ for $r\in(0.67,118.6]$}        &  \multicolumn{1}{p{4.3cm}}{$<0$ for $r\in(0,0.33]$$\cup (118,\infty)$ $>0$ for $r\in(0.33,118]$}  \\ \\
$\rho - p_t$ &  \multicolumn{1}{p{3cm}}{$<0$ for $r\in(0,\infty)$}   &    \multicolumn{1}{p{2.5cm}}{$>0$ for $r\in(0,1]$ $<0$ for $r\in(1,\infty)$}       &   \multicolumn{1}{p{3cm}}{$>0$ for $r\in(0,0.67]$ $<0$ for $r\in(0.67,\infty)$}    \\ \\
$\rho + p_r + 2p_t$ &  \multicolumn{1}{p{3cm}}{$>0$ for $r\in(0,\infty)$ }   &   \multicolumn{1}{p{2.7cm}}{$<0$ for $r\in(0,1.8]$ \hspace{0.5cm} $>0$ for $r\in(1.8,\infty)$}         &    \multicolumn{1}{p{2.7cm}}{$<0$ for $r\in(0,5.4]$ \hspace{0.5cm} $>0$ for $r\in(5.4,\infty)$}   \\ \hline\hline
\end{tabular}
\caption{Summary for results of NFW profile for three different redshift functions}
\label{table:2}
\end{table*}
\subsection{Discussion}\label{subsec3}
In this specific subsection, we will delve into the behavior exhibited by energy conditions within the context of both the PI and NFW models for three different redshift functions. We will thoroughly examine how these models adhere to various energy conditions and elucidate any distinctions or similarities. We have standardized specific free parameters, setting $\alpha=-5,\, r_s=0.5,\, \rho_s=0.02,\, \text{and} \, r_0 = 1$ based on our analysis of shape functions and varying the parameter $\eta$ for the required results. These parameter values were selected to align with the specific requirements of our study and to ensure consistency throughout our analysis.
\begin{itemize}
\item \textbf{First redshift function} $\phi(r)=c$\,\textbf{:} \\
For the specified redshift function, we present graphical representations of the NEC and SEC for different values of $\eta$, particularly $0.05,\, 0.15,\,\text{and} \,0.25$ plotted against the radial coordinate $r$. These representations are depicted in Figs. \ref{fig2} and \ref{fig6}. In these figures, the quantity $\rho+p_r$ shows the violation of the radial NEC at the assumed throat $r_0=1$. Additionally, we have examined the behavior of $\rho+p_t$ for various $\eta$ values and confirmed the satisfaction of the tangential NEC. Moreover, our investigation extends to the SEC, where we observe satisfaction near the throat for different $\eta$ values.
\item \textbf{Second redshift function} $\phi(r)=\frac{1}{r}$\,\textbf{:}\\
In this case, we illustrate graphical depictions of the NEC and SEC across different values of $\eta$: specifically, $0.05$, $0.15$, and $0.25$, plotted against the radial coordinate $r$. These representations are showcased in Figs. \ref{fig3} and \ref{fig7}. Within these visualizations, the quantity $\rho+p_r$ indicates potential violations of the radial NEC, particularly at the assumed throat $r_0=1$. Furthermore, we have analyzed $\rho+p_t$ across various $\eta$ values, confirming adherence to the tangential NEC. Additionally, our investigation extends to the SEC, where we observe violations in the vicinity of the throat for different $\eta$ values.
\item \textbf{Third redshift function} $\phi(r)=\log\left(1+\frac{r_0}{r}\right)$\,\textbf{:}\\
Similarly, in this scenario, we provide graphical representations of the NEC and SEC across various values of $\eta$ as shown in Figs. \ref{fig4} and \ref{fig8}. Remarkably, our observations mirror those obtained with the second redshift function. 

These depictions offer valuable insights into the behavior of these energy conditions within the framework of the given model. For a comprehensive summary of the energy conditions, please refer to Tables \ref{table:1} and \ref{table:2}.
\end{itemize}
\section{Non-Linear $f(Q)$ model}\label{sec5}
\begin{figure*}[t]
\centering
\includegraphics[width=17.5cm,height=6cm]{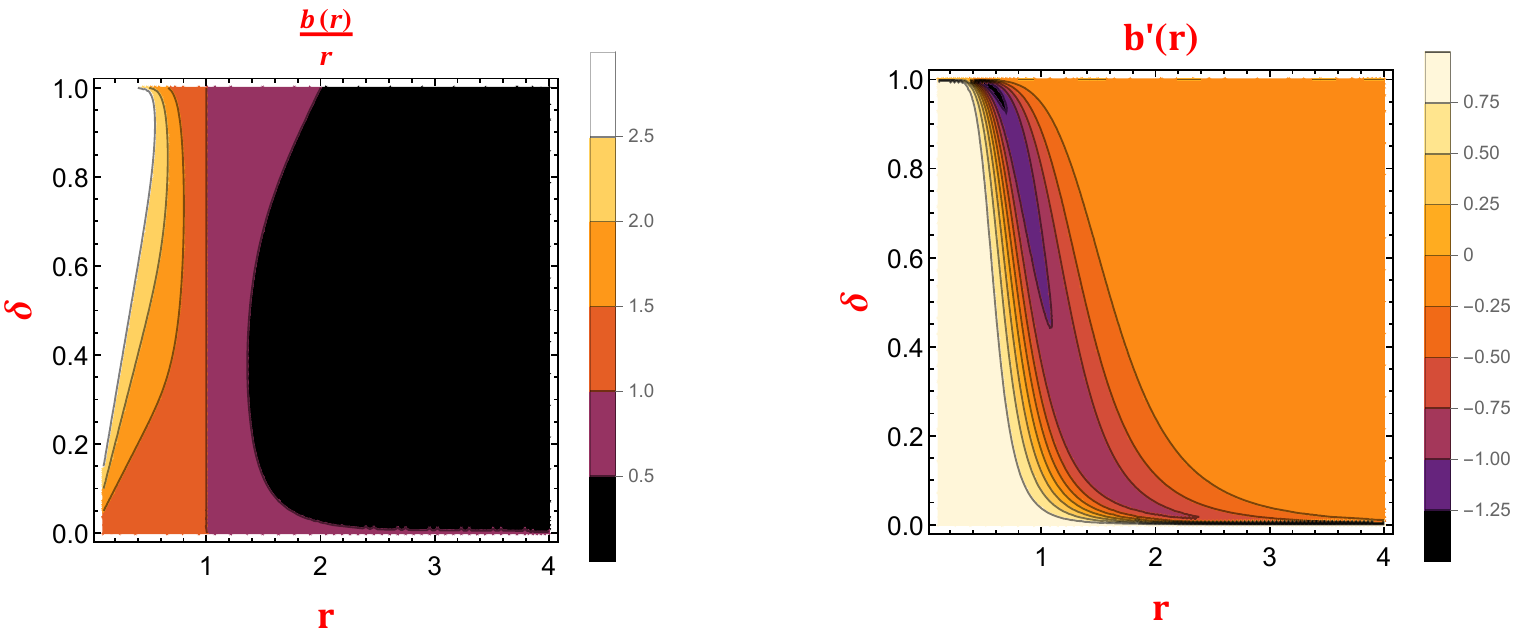}
\caption{The contour plot displays PI profile with the variations in the asymptotic flatness condition \textit{(on the left)} and the flare-out condition \textit{(on the right)} as a function of the radial coordinate `$r$' under the redshift $\phi(r)=\frac{1}{r}$. Additionally, we consider $r_0 = 1$.}
\label{fig11}
\end{figure*}

In this specific section, we leverage the Karmakar condition \cite{M. F. Shamir, G. Mustafa 1} as our guiding principle. Employing a similar methodology, we proceed to derive the shape function applicable to non-constant redshift functions. Specifically, for the redshift function $\phi(r) = \frac{1}{r}$, the corresponding shape function is expressed as follows:
\begin{equation}\label{35}
b(r)=r-\frac{r^5}{\frac{r_0^4 (r_0-\delta ) e^{\frac{2}{r}-\frac{2}{r_0}}}{\delta }+r^4}+\delta ,\;\; 0<\delta<r_{0}\,.
\end{equation}
We adopt a non-linear functional form for $f(Q)$, as articulated in the work \cite{T. Harko,  R. Solanki}. This form is represented as follows:
\begin{equation}\label{37}
f(Q)=Q + m Q^n\,.
\end{equation}
In this formulation, $m$ and $n\neq1$ denote free model parameters, offering flexibility within the model. Consequently, the field equations can be expressed as follows:
\begin{multline}\label{38}
\rho=\frac{-1}{16 \pi  r^6 (r-b)^2}\left(\left(\frac{b \left(-r b'+2 r (b-r) \phi '+b\right)}{r^3 (r-b)}\right)^{n-2}
\right. \\ \left.
\hspace{0.7cm}\times 2 m (n-1) n b  \left(r^2 b \left(r \left(b''-4 \phi '+2 r \phi ''\right)-b' \left(4 r \phi '
\right.\right.\right.\right. \\ \left.\left.\left.\left.
\hspace{0.6cm}+5\right)\right)+r^3 b' \left(b'+2 r \phi '\right)+r b^2 \left(-r \left(b''-8 \phi '+4 r \phi ''\right)
\right.\right.\right. \\ \left.\left.\left.
\hspace{0.6cm}+b' \left(2 r \phi '+3\right)+4\right)+b^3 \left(2 r^2 \phi ''-4 r \phi '-3\right)\right)+r^3 
\right. \\ \left.
\hspace{0.6cm} \times(r-b) \left(b \left(r b'+2 r (b-r) \phi '+b\right)-2 r^2 b'\right)\left(1+m n 
\right.\right. \\ \left.\left.
\hspace{0.6cm} \times \left(\frac{b \left(-r b'+2 r (b-r) \phi '+b\right)}{r^3 (r-b)}\right)^{n-1}\right)-r^3 (r-b) \left(b 
\right.\right. \\ \left.\left.
\hspace{0.6cm} \times\left(-r b'+2 r (b-r) \phi '+b\right)+\left(\frac{b \left(\frac{b-r b'}{r^2-r b}-2 \phi '\right)}{r^2}\right)^n m 
\right.\right. \\ \left.\left.
\hspace{0.6cm} \times r^3 (r-b) \right)+16 \pi  \eta ^2 r^4 (r-b)^2\right)\,,
\end{multline}
\begin{multline}\label{39}
p_r = \frac{1}{16 \pi }\left(\frac{-m n}{b \left(-r b'+2 r (b-r) \phi '+b\right)^2} \left( \left(r^2 b \left(2 r \left((n
\right.\right.\right.\right.\right. \\ \left.\left.\left.\left.\left.
\hspace{0.6cm} -1) \left(b''+2 r \phi ''\right)-4 (n-2) \phi '+14 r \phi '^2\right)+2 b' \left(2 (5
\right.\right.\right.\right.\right. \\ \left.\left.\left.\left.\left.
\hspace{0.6cm} -2 n) r \phi '-5 n+6\right)+b'^2\right)+2 r^3 \left(b'+2 r \phi '\right) \left((n-1)
\right.\right.\right.\right. \\ \left.\left.\left.\left.
\hspace{0.6cm} \times b'-2 r \phi '\right)+2 r b^2 \left(r \left(-(n-1) \left(b''+4 r \phi ''\right)+2 (4 n
\right.\right.\right.\right.\right. \\ \left.\left.\left.\left.\left.
\hspace{0.4cm} -9) \phi '-16 r \phi '^2\right)+b' \left(2 (n-3) r \phi '+3 n-5\right)+4 n-5\right)
\right.\right.\right. \\ \left.\left.\left.
\hspace{0.5cm} +b^3 \left(4 r \left(\phi ' \left(-2 n+3 r \phi '+5\right)+(n-1) r \phi ''\right)-6 n
\right.\right.\right.\right. \\ \left.\left.\left.\left. 
\hspace{0.5cm} +9\right)\right)\left(\frac{b \left(\frac{b-r b'}{r^2-r b}-2 \phi '\right)}{r^2}\right)^n\right)
+\left(\frac{b \left(\frac{b-r b'}{r^2-r b}-2 \phi '\right)}{r^2}\right)^n
\right. \\ \left.
\times m+\frac{2 \left(2 r (b-r) \phi '+b+8 \pi  \eta ^2 r\right)}{r^3}\right)\,,
\end{multline}
\begin{multline}\label{40}
p_t = \frac{1}{16 \pi  r^4 (r-b)^2}\left(\left(1-\frac{b}{r}\right) \left(2 m (n-1) n \phi ' 
\right.\right. \\ \left.\left.
\hspace{0.5cm} \times \left(\frac{b \left(-r b'+2 r (b-r) \phi '+b\right)}{r^3 (r-b)}\right)^{n-2} \left(r^2 b \left(r \left(b'' -4
\right.\right.\right.\right.\right. \\ \left.\left.\left.\left.\left.
\hspace{0.3cm} \times \phi '+2 r \phi ''\right)-b' \left(4 r \phi '+5\right)\right)+r^3 b' \left(b'+2 r \phi '\right) +r b^2
\right.\right.\right. \\ \left.\left.\left.
\hspace{0.3cm} \times\left(-r \left(b''-8 \phi '+4 r \phi ''\right)+b' \left(2 r \phi '+3\right)+4\right)+b^3 \left(2 
\right.\right.\right.\right. \\ \left.\left.\left.\left. 
\hspace{0.3cm} \times r^2  \phi ''-4 r \phi '-3\right)\right)-r^2 \left(\left(b-r b'\right) \left(r (r-b) \phi '+r\right)
\right.\right.\right. \\ \left.\left.\left. 
\hspace{0.3cm} +2 r^2 (r-b)^2 \phi '^2+2\phi ' r (r-b) (r-2 b) +2 r^2 (r-b)^2 \phi ''\right) 
\right.\right. \\ \left.\left.
\hspace{0.4cm} \times \left(m n \left(\frac{b \left(-r b'+2 r (b-r) \phi '+b\right)}{r^3 (r-b)}\right)^{n-1}+1\right)+m r^5 
\right.\right. \\ \left.\left.
\hspace{0.4cm} \times (r-b) \left(\frac{b \left(\frac{b-r b'}{r^2-r b}-2 \phi '\right)}{r^2}\right)^n+r^2 b \left(-r b'+2 r (b-r)
\right.\right.\right. \\ \left.\left.\left. 
\times \phi '+b\right)\right)\right)\,.
\end{multline}
We will now discuss the graphical representation of the derived shape function, as well as the requirements required for the occurrence of a wormhole. To accomplish this, we meticulously select suitable parameters. Initially, our focus lies on examining the behavior of the shape function conditions for the PI profile. The contour plot depicted in Fig. \ref{fig11} showcases both the asymptotic flatness condition and the flaring-out condition with respect to the parameter $\delta$. The left contour in Fig. \ref{fig11} offers insights into the asymptotic behavior of the shape function concerning the parameter $\delta$. Notably, it becomes apparent that as the radial distance increases, the ratio $\frac{b(r)}{r}$ tends towards zero, thus confirming the asymptotic behavior of the shape function. The corresponding graph on the right illustrates the fulfillment of the flaring-out condition at the wormhole throat. Here, we designate the wormhole throat at $r_0=1$.\\
\begin{figure*}[t]
\centering
\includegraphics[width=14.5cm,height=5cm]{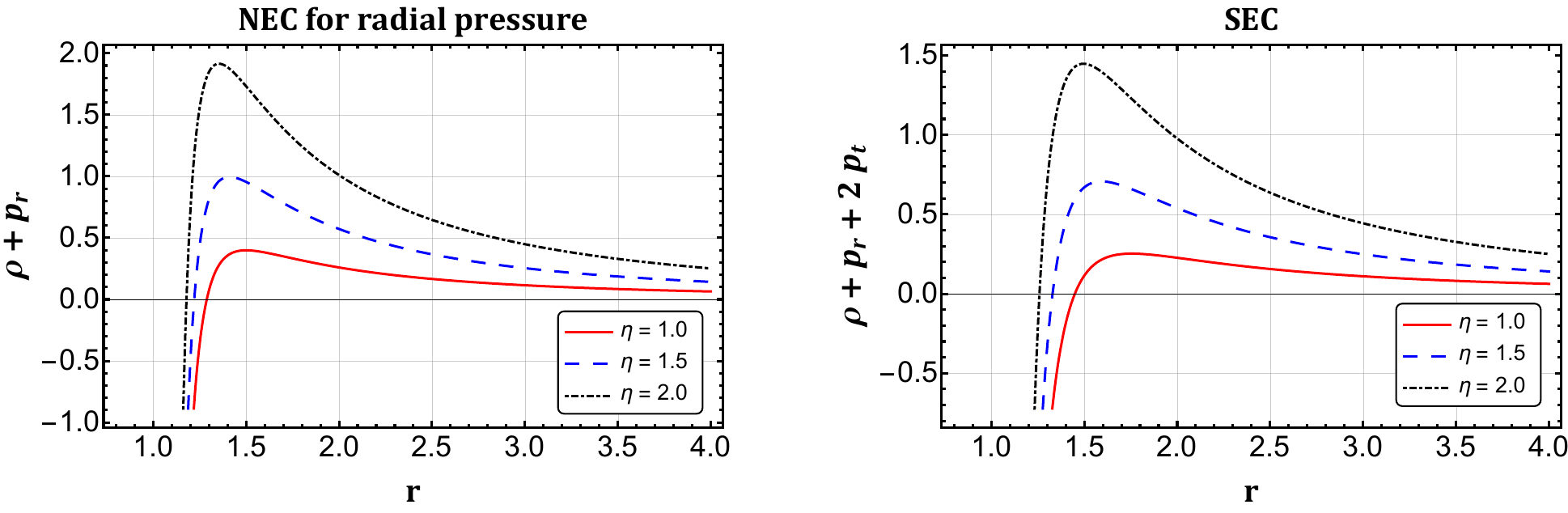}
\caption{The figure displays PI profile with the variations in the NEC for radial pressure and SEC as a function of the radial coordinate `$r$' under the redshift $\phi(r)=\frac{1}{r}$. Furthermore, we keep other parameters fixed at constant values, including $m=2,\,n=2.01,\,\delta=0.5,\, r_s=0.5,\, \rho_s=0.02,\, \text{and} \, r_0 = 1$.}
\label{fig12}
\end{figure*}
Moreover, within this specific context, it becomes evident that the NEC along both radial and tangential directions becomes indeterminate precisely at the wormhole's throat, where $r=r_0$ and can be visualized using Fig. \ref{fig12}. This observation starkly indicates the insufficiency of attaining viable wormhole solutions utilizing the prescribed shape function \eqref{35}. Consequently, we arrive at the conclusion that proposing the non-linear form \eqref{37} proves unsuitable for generating wormhole solutions when paired with the shape function \eqref{35}, applicable to both the PI and NFW profiles.\\
Subsequently, we extend our investigation to alternative non-linear forms such as $Q+\frac{\beta}{Q}$ and $\alpha_1+\beta_1 \log(Q)$, each yielding identical outcomes-namely, the NEC remains undefined precisely at the wormhole's throat for these models as well. Despite these limitations, it is imperative to note the existence of further options for shape functions that we may investigate further.

\section{Amount of exotic matter}\label{sec6}
\begin{figure*}[t]
\centering
\includegraphics[width=14.5cm,height=5.5cm]{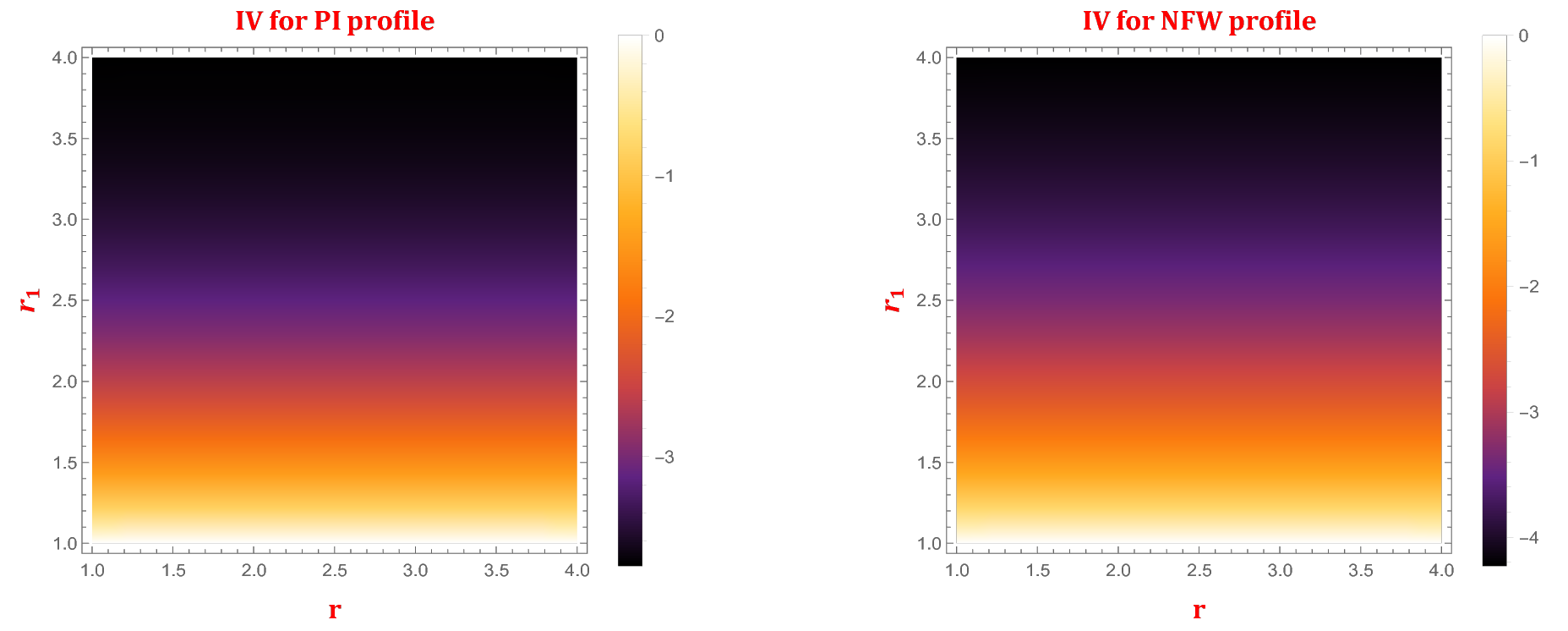}
\caption{The density plot displays the variations in the VIQ under the redshift $\phi(r)=\text{constant}$. Furthermore, we keep other parameters fixed at constant values, including $\alpha=-5,\, r_s=0.5,\, \rho_s=0.02,\, \eta=0.15,\,\text{and} \, r_0 = 1$.}
\label{fig13}
\end{figure*}

\begin{figure*}[t]
\centering
\includegraphics[width=14.5cm,height=5.5cm]{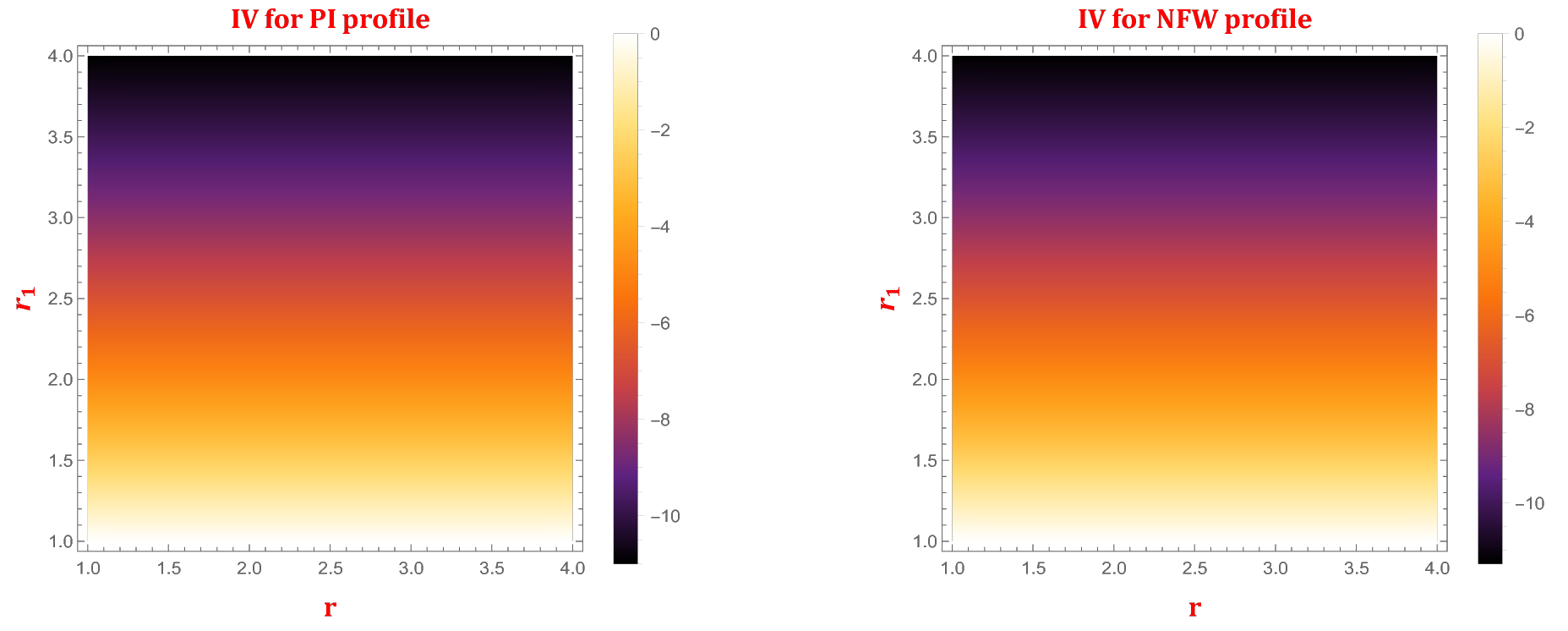}
\caption{The density plot displays the variations in the VIQ under the redshift $\phi(r)=\frac{1}{r}$. Furthermore, we keep other parameters fixed at constant values, including $\alpha=-5,\, r_s=0.5,\, \rho_s=0.02,\, \eta=0.15,\,\text{and} \, r_0 = 1$.}
\label{fig14}
\end{figure*}

\begin{figure*}[t]
\centering
\includegraphics[width=14.5cm,height=5.5cm]{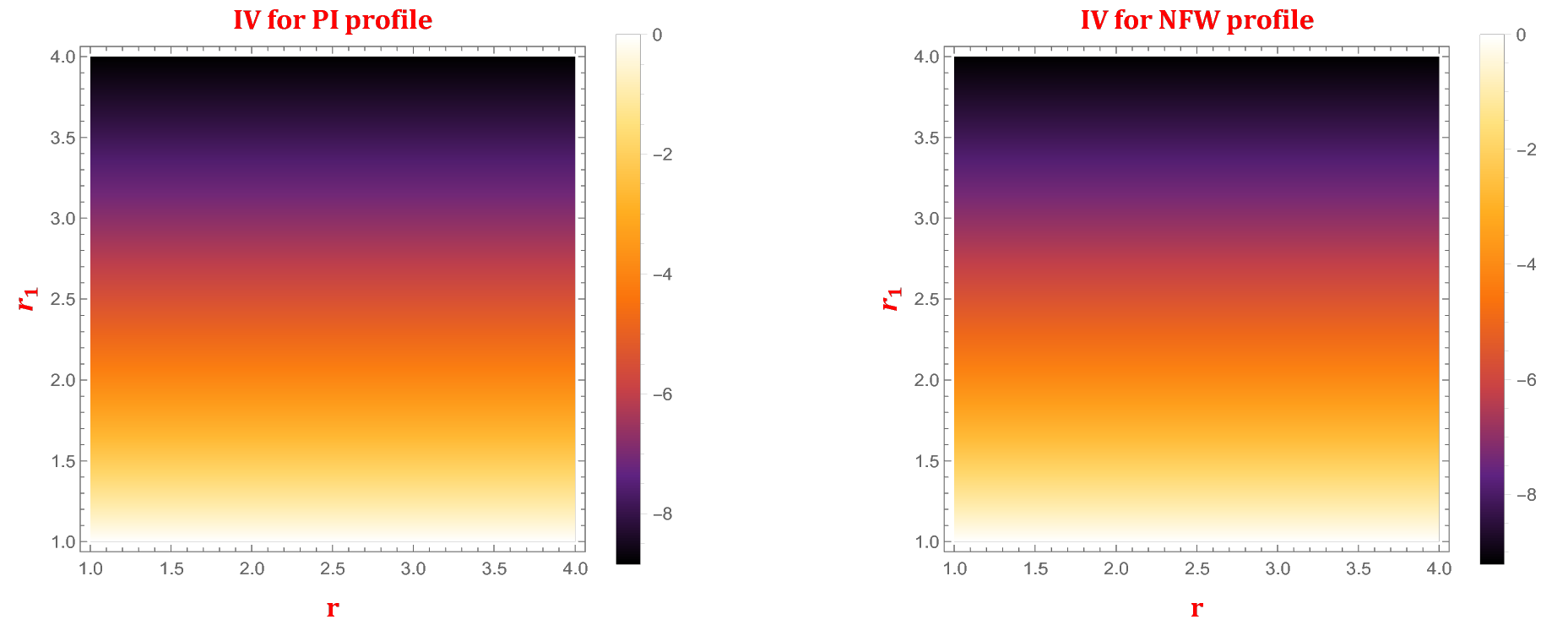}
\caption{The density plot displays the variations in the VIQ under the redshift $\phi(r)=\log\left(1+\frac{r_0}{r}\right)$. Furthermore, we keep other parameters fixed at constant values, including $\alpha=-5,\, r_s=0.5,\, \rho_s=0.02,\, \eta=0.15,\,\text{and} \, r_0 = 1$.}
\label{fig15}
\end{figure*}

Let us now move forward to determine the precise quantity of exotic matter needed to ensure the stability of a wormhole. This task relies on the Volume Integral Quantifier (VIQ) method, first introduced by Visser and his colleagues in their research \cite{M. Visser 2}. This method offers a structured framework for measuring the average amount of matter within the space-time continuum that violates the NEC. The VIQ method can be formally expressed as follows:
\begin{equation}\label{41}
IV=\oint \left[\rho+P_r\right]dV\,.
\end{equation}
The volume can be interpreted as $dV=r^2\,dr\,d\Omega$ with $d\Omega$ the solid angle. Since $\oint dV=2\int_{r_0}^{\infty}dV=8\pi \int_{r_0}^{\infty}r^2dr,$ we have
\begin{equation}\label{42}
IV=8\pi \int_{r_0}^{\infty}(\rho+P_r)r^2dr.
\end{equation}
In the equation presented above, the integration bounds extend indefinitely to infinity. This characteristic has significant implications, as elucidated in the work by F. S. N. Lobo and the collaborators \cite{F.S.N. Lobo}. Their research indicates that for a wormhole to possess asymptotically flat properties, the VIQ, with its bounds extending to infinity over the radial coordinate, should exhibit divergence. To address this, it becomes advantageous to introduce a cut-off scale for the energy-momentum tensor at a designated position denoted as $r_1$. Expanding upon this notion, we propose incorporating an energy-momentum tensor cut-off scale at a specific radial location, $r_1$, within the context of the volume integral formulation for a wormhole. Here, the wormhole is characterized by a field that varies from its throat at $r_0$ to a specified radius $r_1$, ensuring that $r_1$ is greater than or equal to $r_0$. With this condition in mind, we take the volume integral for such a wormhole as follows:
\begin{equation}\label{1122}
IV=8\pi \int_{r_0}^{r_1}(\rho+P_r)r^2dr.
\end{equation}
By leveraging the equation \eqref{1122}, we have conducted an in-depth exploration of the volume integral, analyzing its behavior through visual representations depicted in Density plots \ref{fig13}-\ref{fig15} across various redshift functions. The graphical representations unmistakably indicate that as the parameter $r_1$ converges towards $r_0$, the value of $IV$ progressively tends towards zero. This observation prompts the inference that a relatively minimal amount of exotic matter can serve to stabilize a traversable wormhole. Furthermore, our investigations have revealed that the judicious selection of wormhole geometry can potentially minimize the total quantity of matter violating the Average NEC. These findings underscore the significance of carefully tailored geometries in the context of wormhole stability. For those intrigued by further insights and potential applications of the VIQ, we recommend delving into the works of K. Jusufi \cite{K. Jusufi 2} and O. Sokoliuk \cite{O. Sokoliuk}, which offers additional perspectives and intriguing avenues of exploration.

\section{Conclusions}\label{sec7}
In our present investigation, we have delved into exploring wormhole solutions within the context of $f(Q)$ gravity under the effect of global monopole charge. This comprehensive study entailed utilizing two distinct models for DM halos, juxtaposed against the computed energy density within the framework of $f(Q)$ gravity. Our analysis has generated a set of viable and consistent solutions to the field equations, specifically crafted to describe a Morris-Thorne wormhole (Eq. \eqref{11}) embedded in a DM environment. These models encompass the PI model (Eq. \eqref{19}) and Cold Dark Matter (CDM) halos utilizing the NFW profile (Eq. \eqref{27}) each chosen for its efficacy in modelling the galactic halo. The functional forms of the shape functions have been derived in terms of inverse tangent and other logarithmic functions, showcasing remarkable accuracy within the specified conditions. The salient highlights of our investigation are succinctly outlined as follows:
\begin{enumerate}
    \item The shape functions derived under the PI model and CDM halo with NFW model profiles for the linear form of $f(Q)$ (Eq. \eqref{15}), as illustrated in Figures \ref{fig1} and \ref{fig5}, demonstrate adherence to both the flaring-out condition $b'(r_0)<1$ and asymptotic flatness condition $\frac{b(r)}{r}\to  0$ as $r \to \infty$. These observations underscore the consistency and applicability of the calculated shape functions within these theoretical frameworks, and the observed behavior of the shape functions indicates that the solutions obtained conform to the Morris and Thorne wormhole criterion. This suggests the viability of our solutions within the context of wormhole physics.
    \item From Figures \ref{fig2} and \ref{fig6}, it is evident that the NEC is violated for radial pressure while it is satisfied for tangential pressure. Consequently, both the NEC and WEC are violated overall, which shows that exotic matter may be present at the wormhole throat. Additionally, it is noteworthy that the SEC is fulfilled for both the DM profiles when considering $\phi(r)=\text{constant}$ for various values of $\eta$.
    \item Also, for the cases $\phi(r)=\frac{1}{r}$ and $\phi(r)=\log\left(1+\frac{r_0}{r}\right)$, it is evident from Figures \ref{fig3}-\ref{fig4} and \ref{fig7}-\ref{fig8} that the NEC is violated for radial pressure, whereas it is satisfied for tangential pressure. Consequently, both the NEC and WEC are violated overall, which indicates the possibility of the exotic matter being present at the throat of the wormhole. However, for this case, SEC is violated for both the DM profiles across various values of $\eta$.
    \item Information pertaining to the energy conditions for both DM profiles across varying redshift functions is available in Tables \ref{table:1} and \ref{table:2}. In this context, we assume the location of the wormhole throat to be at $r_0=1$.
    \item The non-linear form of $f(Q)$ (as described in Eq. \eqref{37}) leads to a highly non-linear differential equation. Consequently, it becomes impractical to find an exact analytical solution. In such cases, the Karmakar condition is employed to obtain the shape function (Eq. \eqref{35}), enabling us to navigate the complexities of the differential equation effectively.
    \item For the non-linear Section \ref{sec5}, it becomes evident that the NEC along both radial and tangential directions becomes indeterminate precisely at the wormhole's throat, where $r=r_0$ and can be visualized using Fig. \ref{fig12}.
    \item In Section \ref{sec6}, we used a VIQ parameter to evaluate the quantity of exotic matter necessary to keep the wormhole throat open. It is found that a relatively minimal amount of exotic matter can serve to stabilize a traversable wormhole and can be observed from Figs. \ref{fig13}-\ref{fig15}.
\end{enumerate}
In conclusion, our derived solutions for the PI model and CDM halo with NFW model profiles are deemed physically viable within the framework of $f(Q)$ gravity with global monopole charge and the context of DM halos. \textbf{In the GR context, wormhole solutions within the galactic halo involving monopole charge have been explored in \cite{P. Das}. They considered various dark matter profiles, including NFW, PI, and Universal Rotation Curve (URC) models, confirming that the NFW profile, combined with Global Monopole Charge, provides sufficient support for wormhole structures in the Milky Way galaxy's halo. Additionally, wormhole solutions using NFW, PI, and Thomas-Fermi (TF) dark matter models have been examined in the framework of $f(R)$ gravity \cite{Abdelghani}, demonstrating that the energy conditions, particularly the DEC and SEC, depend on the dark matter models, while the NEC and WEC remain model-independent near the wormhole throat. Furthermore, wormhole solutions with the NFW profile in $f(T)$ gravity have also been studied in \cite{M. Sharif}, while discussions on wormholes surrounded by NFW dark matter within the context of $f(Q)$ gravity have been discussed in \cite{G. Mustafa 2}. However, wormhole solutions incorporating global monopole charge in the context of dark matter have not yet been explored in any modified gravity theories. This gap motivates the study of wormholes with monopole charge supported by dark matter within the $f(Q)$ gravity framework. Our analysis indicates that monopole charges exert a minimal influence on the violation of energy conditions. Future investigations could extend this analysis to $f(R)$ or $f(T)$ gravity to explore the effects of different gravitational models.
}

\section*{Data Availability}
There are no new data associated with this article.

\acknowledgments  MT acknowledges the University Grants Commission (UGC), New Delhi, India, for awarding the National Fellowship for Scheduled Caste Students (UGC-Ref. No.: 201610123801). PKS acknowledges National Board for Higher Mathematics (NBHM) under the Department of Atomic Energy (DAE), Govt. of India, for financial support to carry out the Research project No.: 02011/3/2022 NBHM(R.P.)/R\&D II/2152 Dt.14.02.2022.

\end{document}